\documentclass[aps,preprintnumbers,twocolumn,amsmath,amssymb,prb,showpacs]{revtex4}
\usepackage{bbm}
\usepackage{mathrsfs}
\usepackage{amsmath}
\usepackage{graphicx}
\usepackage{epstopdf}
\usepackage{ulem}
\usepackage{hyperref}

\usepackage{color}

\newcommand{\be}{\begin{equation}}
\newcommand{\ee}{\end{equation}}
\newcommand{\bea}{\begin{eqnarray}}
\newcommand{\eea}{\end{eqnarray}}
\newcommand{\bes}{\begin{split}}
\newcommand{\ees}{\end{split}}

\begin{document}
\title{Deep Learning of Phase Transitions for Quantum Spin Chains
  from Correlation Aspects}
\author{Ming-Chiang Chung$^{1,2,3,4}$}\email{mingchiangha@phys.nchu.edu.tw} 
\author{Guang-Yu Huang$^{2}$}
\author{Ian P.  McCulloch$^{5}$}
\author{Yuan-Hong Tsai$^{6}$}\email{yhong.tsai@gmail.com}
\affiliation{$^1$ Max Planck Institute for  the Physics of Complex
  Systems, N{\"o}thnitze Stra{\ss}e 38, 01187,  Dresden, Germany}
\affiliation{$^2$ Physics Department, National Chung-Hsing University, Taichung, 40227, Taiwan}
\affiliation{$^3$ National Center for Theoretical Sciences, Physics
  Divison, Taipei,  10617, Taiwan  }
\affiliation{$^4$ Physics Department, Northeastern university, 
360 Huntington Ave., Boston, Massachusetts 02115, U.S.A. }
\affiliation{$^5$ School of Mathematics and Physics, the University of Queensland, St. Lucia, QLD 4027, Australia} 
\affiliation{$^6$ AI Foundation, Taipei 106, Taiwan}

\begin{abstract}
Using machine learning (ML) to recognize different phases of matter and to infer the entire phase diagram has proven to be an effective tool given a large dataset. In our previous proposals, we have successfully explored phase transitions for topological phases of matter at low dimensions either in a supervised or an unsupervised learning protocol with the assistance of quantum information related quantities. In this work, we adopt our previous ML procedures to study quantum phase transitions of magnetism systems such as the XY and XXZ spin chains by using spin-spin correlation functions as the input data. We find that our proposed approach not only maps out the phase diagrams with accurate phase boundaries, but also indicates some new features that have not been observed in the field of machine learning before. In particular, we define so-called relevant correlation functions to some corresponding phases that can always distinguish between those and their neighbors. Based on the unsupervised learning protocol we proposed [Phys. Rev. B 104, 165108 (2021)]\cite{ChungTsai1DUSL}, the reduced latent representations of the inputs combined with the clustering algorithm show the connectedness or disconnectedness between neighboring clusters (phases), just corresponding to the continuous or disrupt quantum phase transition, respectively. This property reminds us of the behavior of order parameters. Moreover, in the Silhouette analysis we show that the ferromagnetic states in the XXZ model with various anisotropy parameters correspond to almost the same Silhouette value, while the critical or anti-ferromagnetic states behave quite differently. The analysis further indicates that the minima of Silhouette values are close to the phase transition points, showing strong positive correlation. These results again justify the usefulness of our proposed ML procedures and move a step toward understanding the relation between ML and quantum phase transitions from correlation function aspects.
\end{abstract} 
\pacs{}

\date{\today}
\maketitle

\section{Introduction} 
Quantum phase transitions (QPTs) have intrigued people for several decades\cite{Sachdev}. Different from classical phase transitions accessed by varying temperatures, QPTs are driven by altering a non-thermal, physical parameter at zero temperature. While Ginzburg-Landau theory\cite{GinzburgLandau} is often employed to describe phase transitions, predicting QPTs in a one-dimensional (1D) quantum system with continuous symmetry and sufficient short-range interactions brings challenges to it. It is because the presence of a local order parameter in such a system would violate the Mermin-Wagner theorem\cite{MerminWagner}, and thus people look for effective alternatives. Computing correlation functions is just one of the alternatives to detect the long-range orders of the quantum phases and predict QPTs.

Several numerical techniques, such as exact diagonalization\cite{ED}, quantum Monte Carlo simulations\cite{QMC}, and the density matrix renormalization group (DMRG)\cite{DMRGBook,Schollwoeck}, can be used to compute correlations and ground-state wave functions. However, these methods are often computationally heavy and have limitations. This makes them difficult to map out the entire phase diagram in the parameter space. Recently, machine learning (ML) has gained significant attention not only in computer science, but also in physics for its ability to reveal hidden structures of correlations, entangled quantities, and complex wave functions. Moreover, ML is becoming a powerful tool for scientific researchers, as it is completely data-driven. For instance, given a set of data, a neural network (NN) can be trained to identify patterns or relate to condensed representations (such as class labels) for the data. The trained model can then predict unseen data points. This advantage brings the physics community's interest in using ML to determine the phase boundaries between different phases of matter, including quantum phase transitions\cite{Ohtsuki16,Nieuwenburg17,Carrasquilla17,Broecker17,Wetzel17,kim17a,kim17b,Zhang18,Sun18,Carvalho18,Ming19,Caio19,Dong19,Zhang19,Zhang21,ChungTsai1DSL, ChungTsai2DSL,Uria-Alvarez22,wang16,Tanaka17,Hu17,Broecker17b,Wetzel17b,Chng18,Liu18,Durr19,Scheurer19,Kottmann20,Che20,Lidiak20,Yu20,Greplova20,Scheurer20,Dawid20,Kaming21,ChungTsai1DUSL,Magnifico22,Contessi22}. 

There are two main approaches to using ML to classify phases of matter. The first approach is supervised learning, where the training data are labeled with known regimes\cite{Ohtsuki16,Nieuwenburg17,Carrasquilla17,Broecker17,Wetzel17,kim17a,kim17b,Zhang18,Sun18,Carvalho18,Ming19,Caio19,Dong19,Zhang19,Zhang21,ChungTsai1DSL, ChungTsai2DSL,Uria-Alvarez22}. This way is more accurate at identifying phase boundaries, but it requires prior knowledge of the system, such as the number of phases in a particular parameter range. The second approach is unsupervised learning, which does not require any prior knowledge or labeling and instead learns from the training data itself\cite{wang16,Tanaka17,Hu17,Broecker17b,Wetzel17b,Chng18,Liu18,Durr19,Scheurer19,Kottmann20,Che20,Lidiak20,Yu20,Greplova20,Scheurer20,Dawid20,Kaming21,ChungTsai1DUSL,Magnifico22,Contessi22}. While this approach is more difficult and demanding, it is a natural option when exploring a parameter space for scientific discovery without prior knowledge.
 
In our previous works, we proposed a protocol for locating phase boundaries using machine learning, with either supervised\cite{ChungTsai1DSL, ChungTsai2DSL} or unsupervised learning\cite{ChungTsai1DUSL}. In particular, under unsupervised learning, we fed input data into an autoencoder\cite{Lecun87,Kamp88,Hinton94} to extract effective features of data and then applied principal component analysis (PCA)\cite{Pearson01,Jolliffe02} to determine the necessary feature dimensions. We next used the K-means clustering algorithm\cite{MacQueen67,Lloyd82} followed by Silhouette analysis\cite{Rousseeuw87,Hennig15} to determine the total number of phases without prior knowledge. Finally, we used supervised learning to improve the precision of the phase boundaries by taking the most confident points in Silhouette analysis as labeled training seeds. Our method was successful in finding topological phase boundaries using quantum-information-related quantities as the input data.

As mentioned before, correlation functions are often natural quantities to detect long-range orders in QPTs. However, they are usually not computationally ``cheap'', with hidden structures, and can be inaccurate near critical regimes. Therefore, in this study, we employ the ML approach for identifying QPTs, assisted by using correlation functions as the input data type instead of quantum-information-related quantities. We benchmark our approach on two classic 1D magnetism systems, namely, the XY and XXZ models. The correlation functions of the 1D XY model can be calculated by fermionization of the spin system\cite{LiebSchultzMattis,BarouchMcCoy}, whereas those of the XXZ model can be calculated by DMRG\cite{White,DMRGBook,Schollwoeck}.  

Besides showing our proposed ML protocol effective for detecting QPTs, a few essential observations are highlighted as follows. First, there are two types of phase transitions: First-order (disrupt) and second-order (continuous). It is intriguing if the proposed protocol of ML can distinguish these two. In the XY model, there is a second-order phase transition, while in the XXZ model, there are both of them. In unsupervised learning, the reduced latent representations of the input data after Silhouette analysis show huge differences between the two types of phase transitions, unveiling an important signature to distinguish between them when using ML. 

Second, one would like to know how the precision of correlation functions influences the accuracy of the predicted phase transition points (TPs). As to the XY model, since all the numerical calculations can be done with great precision, the predicted phase boundaries are of the order of the statistical error for ML, suggesting that precise phase boundaries can be achieved. On the other hand, for the XXZ model, the accuracies of the correlation functions may fluctuate as a function of the anisotropy of the model in the z-direction with some fixed truncation basis. Therefore, the accurate positions of the predicted phase boundaries also depend on whether they are first-order or second-order phase transitions. For the former case, the accuracy is not influenced much, while for the latter one, due to the slow gap opening, the predicted results could deviate more. It turns out that for achieving better accuracy, more delicate treatments should be made when training our machines. 

The final one is about the relevancy of the correlation functions in the corresponding phases. We find that if a correlation function is relevant in a certain phase, it can be naturally used to determine the phase boundary between that one with the other. On the other hand, without any relevancy of the correlation function in a phase, it could easily fail to distinguish between the phases of matter unless the machine could find other patterns to recognize them.

In this paper, we first introduce the XY and XXZ models and their spin-spin correlation functions. Then we show the procedures and the results of ML under supervision for recognizing different phases. This is used to show the effectiveness of taking correlations as the input data and can be compared with the later unsupervised method. In the next section, the protocol and the results of unsupervised learning are discussed. At the end we  conclude with several important discussions about the accuracies of the predicted phase boundaries and the signatures of the first order and second order phase transitions in ML approach. Moreover, why the relevancy of the correlation functions for the corresponding phases are essential to make the ML approach effective are also discussed.

\section{XY and XXZ models} 
\begin{figure}[th]
	\begin{center}
		\includegraphics[width=7.5cm]{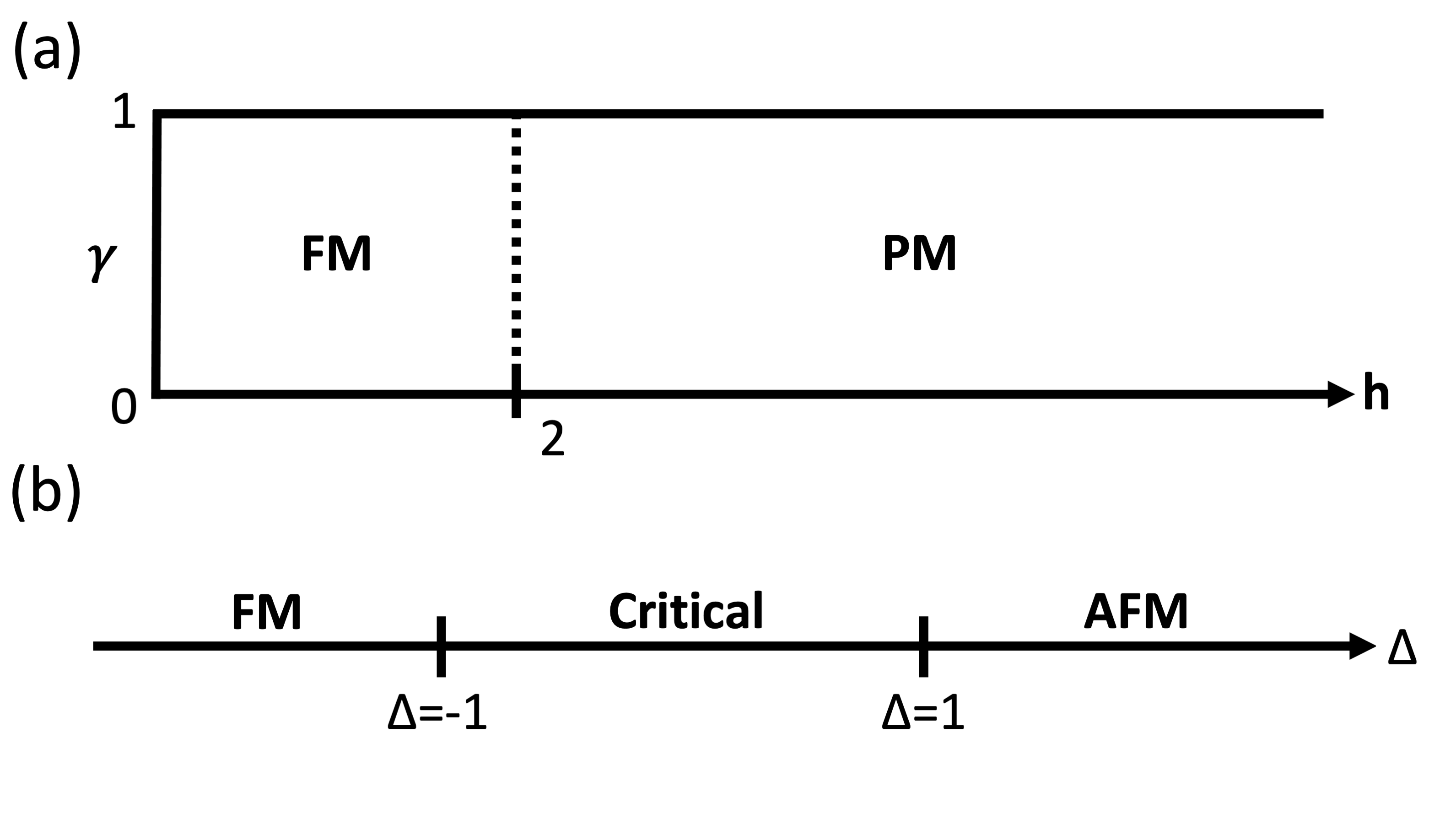}
		\caption{ (a) Phase diagram of the 1D XY model. For $h<2$, the system is in the ferromagnetic (FM) state, whereas in the case of $h>2$, the system is paramagnetic. (b) Phase diagram of the 1D XXZ model. In the case of $\Delta > 1$, the system is in the anti-ferromagnetic (AFM) state, while for $\Delta < -1$, it is FM. In between ($-1<\Delta <1$), it is in the gapless critical state.}
		\label{fig:PD}
	\end{center}
\end{figure}

\subsection{XY model} 
The Hamiltonian of XY model reads
\begin{equation} \label{XYmodel} 
   H_{XY} = -\frac{1}{2} \sum_{j=1}^{N} [(1+\gamma) \sigma_j^{x} \sigma_{j+1}^{x} +(1-\gamma)
   \sigma_j^{y} \sigma_{j+1}^y] -\frac{h}{2}\sum_{j=1}^N  \sigma_j^{z} ,   
\end{equation} 
describing a chain of $N$ spin-1/2's  interacting ferromagnetically with their
nearest neighbors. Here $\sigma^{a}_i$ with $a=x,y,z$ at site $i$ are Pauli matrices, obeying the usual commutation relations, $[\sigma^{a}_i, \sigma^{b}_j]=2i\delta_{ij}\epsilon^{abc}\sigma^{c}_i$. The Hamiltonian is symmetric under the mapping $\gamma \rightarrow -\gamma$ or $h \rightarrow -h$, therefore one can only consider the cases where $\gamma \ge 0 $ and $h \ge 0$ without loss of
generality. For $\gamma =1$, one obtains the 1D transverse field Ising model. In our study, we stress the cases with stronger spin-spin interaction along the $x$-direction, and consider only the cases with
$0 \le \gamma \le 1$.  

The solution of XY model in one dimension has been derived for a long time since Lieb, Schultz and Mattis found the exact solution through the Jordan-Wigner transformation\cite{LiebSchultzMattis}. With the definition, $\sigma_j^{+} = \frac{1}{2}(\sigma_j^x +i \sigma_{j}^{y}) $ and $\sigma_j^{-} = \frac{1}{2}(\sigma_j^{x} - i \sigma_j^{y} $), the Jordan-Wigner transformation in terms of spinless Fermion operators $c_j$ and $c^{\dagger}_j$ is defined as follows:
\begin{eqnarray} \label{JW}
\sigma^{-}_j & = &\exp{\left(- i\pi \sum_{i=1}^{j-1} c_i^{\dagger}
                         c_i\right)}  c_j,  \nonumber \\
\sigma^{+}_j & = & c_j^{\dagger} \exp{\left(i\pi \sum_{i=1}^{j-1}
                   c_i^{\dagger} c_i\right) } . 
\end{eqnarray}
The Hamiltonian is then transformed to the form:
\begin{equation}
  H_{XY} = - \left[ \sum_{j=1}^N (c_j^{\dagger} c_{j+1} + \gamma
  c_j^{\dagger} c_{j+1}^{\dagger} + \mathrm{H.c.} ) +h c_j^{\dagger}
  c_j  \right] + \frac{Nh}{2} .
\end{equation}
Under the periodic boundary ($c_j = c_{N+j}$) condition the Hamiltonian can be further Fourier
transformed into
\begin{eqnarray}
      H_{XY} &=&  -\sum_{k}  2(\cos{k} + h/2) c_k^{\dagger} c_k  \nonumber  \\ 
       &-& i \gamma \sin{k} (c_k^{\dagger} c_{-k}^{\dagger} + c_k c_{-k}) + \frac{N}{2} h, 
\end{eqnarray}
 where the lattice constant is set to one. The system is decomposed into ``noninteracting'' (diagonalizable) momentum subspaces, and thus the interested correlation functions can be calculated within the noninteracting subspaces. 

\begin{figure}[th]
	\begin{center}
		\includegraphics[width=7.5cm]{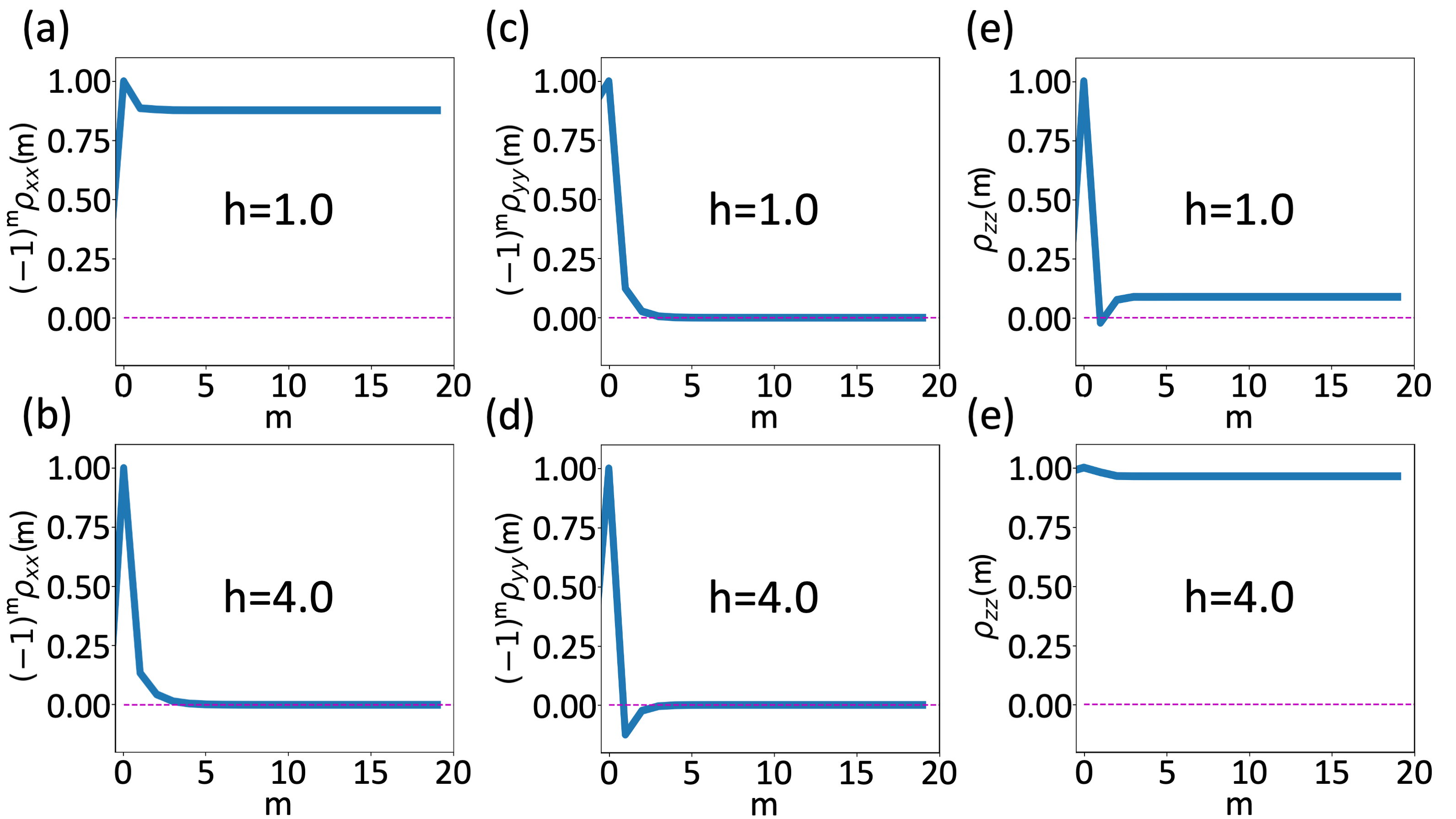}
		\caption{Spin-spin correlation functions for the XY
                  model with various $h$. (a) $(-1)^n\rho_{xx}(m) $ for $h=1.0$. (b)
                  $(-1)^m \rho_{xx}(m)$
                  for $h=4$. (c) $(-1)^m \rho_{yy}(m) $ for $h=1.0$. (d) $(-1)^m\rho_{yy}(m)$
                  for $h=4$. (e) $\rho_{zz}$ for $h=1$. (f) $\rho_{zz}$
                  for $h=4$. For all subfigures, $\gamma=0.5$. One can see that $\rho_{xx}$ is a
                  relevant correlation function in the ferromagnetic
                  phase, whereas $\rho_{zz}$ is relevant in the
                  paramagnetic phase. 
                }
		\label{fig:SSCFXY}
	\end{center}
\end{figure}

The essential spin-spin correlation functions we considered here are defined as 
\begin{equation}
 \rho_{a a}(m) = \langle \Psi_0 \mid \sigma_0^{a} 
 \sigma_m^{a} \mid \Psi_0 \rangle, 
\end{equation}  
where $a = x,y,z$ and $\Psi_0$ is the ground state of the
XY model (\ref{XYmodel}). Note that $\rho_{a a}$ depends only on the relative distance, $r_m-r_0$, between the sites $0$ and $m$ due to the translational invariant property.  Using Jordan-Wigner transformation, Lieb, Schultz and Mattis\cite{LiebSchultzMattis} showed that $\rho_{xx}(m) $ can be transformed to the Toeplitz determinant
\begin{equation} \label{rhoxx}
\rho_{xx}(m) =  
\begin{vmatrix} 
G_{-1} & G_{-2} & \cdots &G_{-m} \\
G_0    &G_{-1} & \cdots  &G_{-m+1} \\
\cdot & \cdot & \cdots & \cdot \\
 \cdot & \cdot & \cdots & \cdot \\
 \cdot & \cdot & \cdots & \cdot \\
G_{m-2} & G_{m-3} & \cdots & G_{-1} 
\end{vmatrix},
\end{equation}
where $G_{l-n} = \langle \Psi_0 \mid  B_l A_n \mid \Psi_0 \rangle$
with the definitions: $A_n = c_n^{\dagger} +c_n$ and $B_l =
c_l^{\dagger} +c_l$. While $\rho_{yy}(m)$ has the same structure as $\rho_{xx}(m)$,
\begin{equation} \label{rhoyy}
\rho_{yy}(m) =  
\begin{vmatrix} 
G_{1} & G_{0} & \cdots &G_{-m+2} \\
G_2    &G_{1} & \cdots  &G_{-m+3} \\
\cdot & \cdot & \cdots & \cdot \\
 \cdot & \cdot & \cdots & \cdot \\
 \cdot & \cdot & \cdots & \cdot \\
G_{m} & G_{m-1} & \cdots & G_{1} 
\end{vmatrix},
\end{equation}
$\rho_{zz}(m)$ has a much simpler form, 
\begin{equation} \label{rhozz}
\rho_{zz}(m) = 
\begin{vmatrix}
G_o & G_m \\
G_{-m} & G_0
\end{vmatrix},
\end{equation}
due to the fact that spin-1/2 $S_i^z=\frac{1}{2}\sigma_i^z$ is a local operator unlike $S_i^x$ or $S_i^y$.
Through the properties of the ground state $|\Psi_0\rangle$ and the advantage from the periodic boundary condition, the correlation function $G_{l-n}$ can be calculated by
\begin{equation} \label{BA}
G_{l-n}  = \frac{1}{\pi} \int_{0}^{\pi} dk\frac{\gamma \sin{k} \sin{kR} - (\cos{k} +h/2) \cos{kR}}{\sqrt{(\cos{k}+h/2)^2+\gamma^2\sin^2{k}}}, 
\end{equation} 
where $R=r_l-r_n$, indicating the distance between sites $l$ and $n$. Interested readers are referred to Appendix A for more details.

According to Mermin-Wagner theorem\cite{MerminWagner}, continuous symmetry breaking does not occur spontaneously in 1D quantum systems at zero temperature. In other words,  no local order parameter can gain any finite expectation value for such systems. However, the phase transitions can still occur and be observed by using appropriate correlation functions. For instance, in Fig.~\ref{fig:PD}(a), the phase diagram is shown for the XY model, and the spin-spin correlation functions for the XY model are shown in Fig.~\ref{fig:SSCFXY}.  When $h<2 $, the system is in an ordered phase, whose correlation $\rho_{xx}$ asymptotically approaches to a constant as $m\rightarrow \infty$. This property represents the
long-range order of a ferromagnetic (FM) phase. On the other hand, when $h>2$, a
disordered phase is observed due to the asymptotic behavior $\rho_{xx}
\rightarrow 0$ and $\rho_{zz} \rightarrow 1$. Additionally, these properties indicate that the disordered phase is actually paramagnetic (PM).  Note that all three correlation functions $\rho_{xx}$, $\rho_{yy}$ and $\rho_{zz}$ behave quite differently in FM and PM phases. Therefore, they can serve as good candidates for the input data format of deep learning machines to distinguish different quantum phases. 


\subsection{XXZ model}

\begin{figure}[th]
	\begin{center}
		\includegraphics[width=7.5cm]{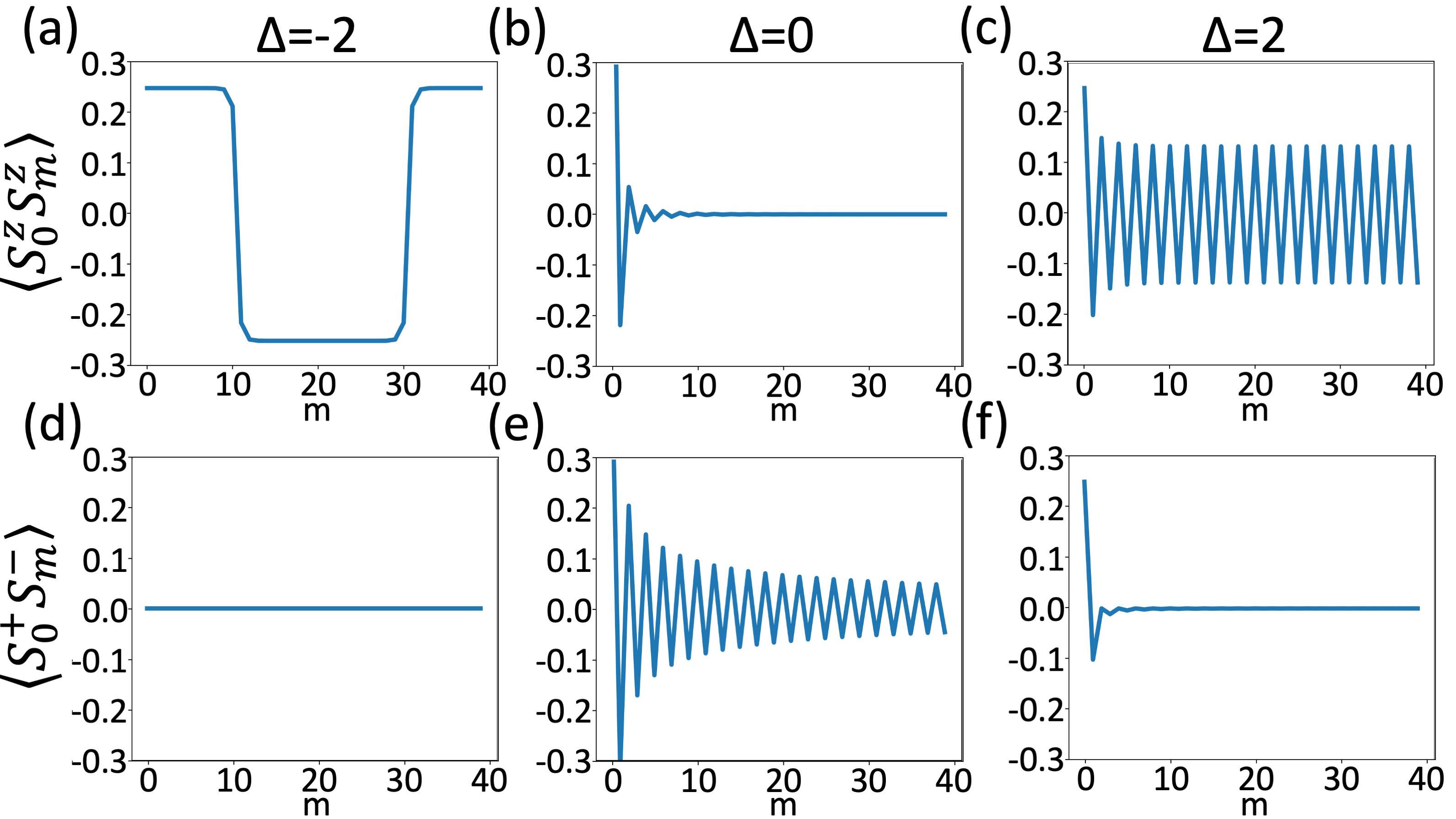}
		\caption{Spin-spin correlation functions for the XXZ model with various $\Delta$. (a), (b) and (c) are $\langle S_0^z S_m^z \rangle$ for $\Delta=-2$ (FM), $\Delta=0$ (critical) and $\Delta=2$ (AFM), respectively. (d), (e) and (f) are $\langle S_0^{+} S_m^{-}\rangle$ for $\Delta=-2$ (FM), $\Delta=0$ (critical) and $\Delta=2$ (AFM), respectively. We can see that $\langle S_0^z S_m^z\rangle$ is relevant in the AF and FM phases, while $\langle S_0^{+} S_m^{-}\rangle$ is only relevant in the critical phase. 
                }
		\label{fig:SSCFXXZ}
	\end{center}
\end{figure}

1D XXZ model with zero magnetic field, served as the simplest model for explaining strongly correlated systems, has been ``the'' model to investigate the
magnetism and physics of some other related 1D interacting  fermion systems since Werner Heisenberg proposed XXX model \cite{HeisenbergXXX} and Hans Bethe gave the first solution by using Bethe Ansatz \cite{Bethe}. The Hamiltonian of XXZ model has the following form
\begin{equation}
H_{XXZ} = \sum_j S_j^x S_{j+1}^x +   S_j^y S_{j+1}^y + \Delta  S_j^z S_{j+1}^z, 
\end{equation}
where $S^{a} = \sigma^{a}/2$. Its phase diagram is shown in Fig.~\ref{fig:PD}(b). For $\Delta <-1$, the system is in the FM state,
while for $\Delta >1$, it becomes AFM. In between, {\it i.e.}, in the case that $-1<\Delta<1$, it is in the XY phase (critical regime) due to the fact that its corresponding correlation length is divergent. Note that for $\Delta =1$ the model corresponds to the AFM Heisenberg model (XXX model), whereas for $\Delta = -1$ it is the FM Heisenberg model.

The XXZ model is a quantum integrable model, which means that they can be solved by using Bethe Ansatz (BA) \cite{Bethe}. However, it has always been a hard topic to obtain spin-spin correlation functions with BA.  For instance, to calculate $\langle S_0^{+}S_m^{-}\rangle$, it involves several complicated multi-dimensional integrals with the highest dimension equal to $2m+1$  in the complex plane. In other words, for $m\ge 2$ it becomes numerically difficult to calculate the spin-spin correlations \cite{Kitanine}. 

The correlation functions for the critical phase, where $-1<\Delta<1$,  can be calculated via bosonization approach \cite{Giamarchi}, which only considers the linear energy spectra near two Fermi points, and maps the Hamiltonian into an effective bosonic model (Luttinger liquid) parametrized by the group velocity and Luttinger parameters \cite{Giamarchi}. The outcome is that in the thermodynamic limit, where the correlation length of the system is infinite, the spatial correlations decay algebraically with powers determined by the Luttinger parameters and behave like a (quasi) long-range order. However, this method only works in the massless Luttinger regime. For the other massive phases, {\it e.g.}, FM and AFM, Luttinger liquid analysis fails. For our purpose, we need to calculate correlations in all regimes, and hence some other alternative methods must be adopted.

We overcome the issue mentioned above by leveraging a numerical method called density-matrix renormalization group (DMRG),  which was developed by Steven White in 1992\cite{White, DMRGBook}. The low energy spectrum, the correlations and other useful physics properties of 1D systems can be calculated efficiently by DMRG due to the low bipartite entanglement of 1D systems. In particular, the calculations can be quick and precise with very low truncation errors. Figs.~\ref{fig:SSCFXXZ}(a), (b)  and (c) show  $\langle S_0^{z} S_m^{z}\rangle$ for $\Delta \le -1$, $-1 < \Delta \le 1$ and $\Delta>1$, respectively, using a specific algorithm called iDMRG with the length of unit cell $L_{UC} =40$ and the bond dimension $m_b=100$\cite{Ian1}, enforcing $U(1)$ symmetry in the $S_z=0$ sector.  The iDMRG algorithm produces Matrix Product State wavefunctions that are infinite in size, and translationally invariant under shifts of the $40$ site unit cell.  The large unit cell allows the system to partially phase-separate in the ferromagnetic region, as shown in Fig.\ref{fig:SSCFXXZ}(a), whereas in the critical phase the wavefunction will be translationally invariant under a $2$-site shift (which is the minimum possible translational invariance allowed by the $U(1)$ symmetry for spin-$1/2$ systems). One can see that for Fig.~\ref{fig:SSCFXXZ}(a) the correlations develop some plateaus at the values $1/4$ and $-1/4$, showing FM property with total magnetization equal to zero due to the $U(1)$ symmetry. In the critical regime,  $-1<\Delta \le 1$,  $\langle S_0^{z} S_m^{z}\rangle$ quickly decays to zero as a function of the distance, which indicates that there is no long-range order at all, as shown in Fig.~\ref{fig:SSCFXXZ}(b). For the AFM ($\Delta>1$),  $\langle S_0^{z} S_m^{z}\rangle$ has a zig-zag curve, which represents the AFM long-range order, {\it i.e.}, $(-1)^m \langle S_0^{z} S_m^{z}\rangle$ decays to a constant, as shown in Fig.3(c). Therefore, $S_0^z S_m^z$ can be viewed as a relevant operator for the FM and AFM phases, whereas for the critical phase it is irrelevant. 

On the other hand, the behaviors of $\langle S_0^{+} S_m^{-}\rangle$ are shown in Figs. 3(d), (e), and (f). 
From the figures one can see that $\langle S_0^{+} S_m^{-}\rangle$ decays to zero for both FM and AFM phases, while for critical phase it shows the XX long-range order similar to the XY model for $\gamma =1$. Therefore, on the contrary to $S_0^z S_m^z$,  $S_0^{+} S_m^{-}$ can be viewed as a relevant operator for the critical phase, whereas for both FM and AFM it is irrelevant. The relevancy of the correlation operators are very important for the outcomes of deep learning process, as we will discuss about it later. 

\section{Supervised  Learning}
\subsection{Data Preparation} 
In the previous studies \cite{ChungTsai1DSL, ChungTsai2DSL, ChungTsai1DUSL} we used quantum information related quantities, such as Majorana correlation matrices (MCMs), block correlation matrices (BCMs), one-particle entanglement spectra (OPES), or one-particle entanglement eigenvectors (OPEEs) as input data for deep learning processes. The whole data construction of the entanglement-related data is based on the calculations of the fermion-fermion correlation functions, and thus we here take the spin-spin correlation functions of the XY and XXZ models in analogy with those of fermion models.

For the XY model, we prepare each of $\rho_{xx}(r_j-r_i)$, $\rho_{yy}(r_j-r_i)$ and $\rho_{zz}(r_j-r_i)$ as a $m_s\times m_s$ matrix, where $r_i, r_j = 0,1,..., m_s-1$, by calculating $G_{l-n}$ and the Toeplitz determinant for each entry. The subsystem length, $m_s=\text{max }|r_j-r_i|+1$, is set to be less than $40$ and the precision of the integrals is $10^{-16}$ by using Romberg integrals.  

On the other hand, for the XXZ model, we use iDMRG algorithm with matrix product states (Matrix Product Toolkit by Ian P. McColluch\cite{Ian2}) to compute correlation functions, $\langle S_i^{\alpha} S_j^{\alpha^\prime}\rangle$, where $\alpha,\alpha^\prime=z,+,-$.  We take the length of the unit cell $L_{UC}=40$, 
and the bond dimension $m_b = 100$. For the ground state calculation, its energy is compared with the result of Bethe Ansatz and the precision is distributed within the range of $10^{-12}$ to $10^{-4}$. For the critical phase the accuracy is relatively lower, whereas for the gapped phases (FM and AFM) the precision is much higher.  

\subsection{Deep Learning Processes}

\begin{figure}[th] 
	\begin{center}
		\includegraphics[width=9cm]{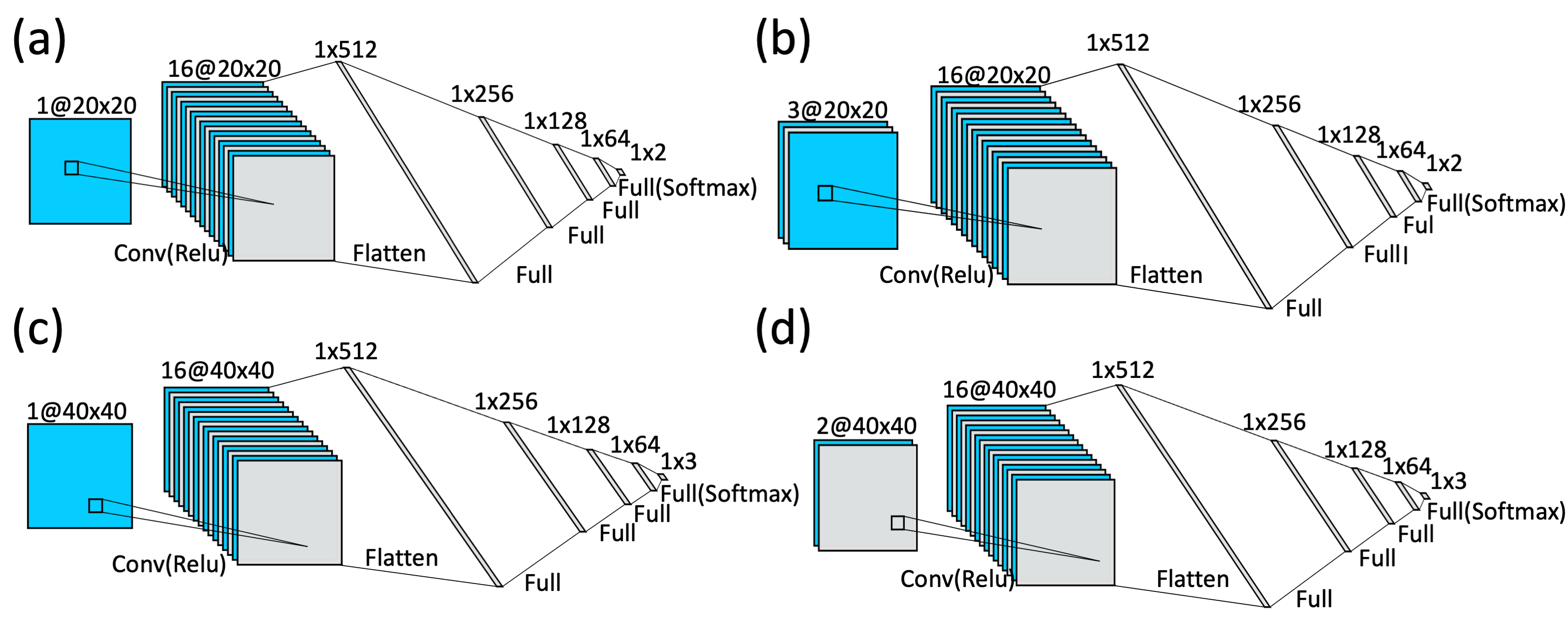}
		\caption{The schematic illustrations of the convolutional neural networks for various input data: (a) $\rho_{xx}$, $\rho_{yy}$ or $\rho_{zz}$, (b) $\rho_{xx}$, $\rho_{yy}$ and $\rho_{zz}$ all together, (c) $\langle S_0^z S_m^z \rangle$ or $\langle S_0^{+} S_m^{-}\rangle$, (d) both $\langle S_0^z S_m^z \rangle$ and  $\langle S_0^{+} S_m^{-}\rangle$. 
                }
		\label{fig:CNN}
	\end{center}
\end{figure}

The usage of different types of the neural network architectures to recover phase transitions often depends on data types and their characteristics. For instance, the types like matrices or tensors can be naturally viewed as single- or multi-channel 2D ``images'' and then a convolutional neural network (CNN) would be an efficient candidate for the pattern recognition in the ``images'' due to its inductive priors such as translational invariance and the spatial locality. Here, the matrices based on the spin-spin correlation functions just belong to this kind of data types.

In the XY model there are three essential correlation functions: $\rho_{xx}, \rho_{yy}$ and $\rho_{zz}$. One can choose one of them as the input data or put all of them into the machines with model architectures shown in Figs.~\ref{fig:CNN} (a) and (b), respectively. In the situation (a), each ``image'' contains information of  a $m_s\times m_s$ matrix, whereas in (b), the input data are ``images'' of dimension, $3\times m_s\times m_s$. `3' indicates the dimension along which all $\rho_{xx}, \rho_{yy}$ and $\rho_{zz}$ are stacked together. On the other hand, for the XXZ model, there are two important correlations: $\langle S_0^{z} S_m^z \rangle$ and $\langle S_0^{+} S_m^{-} \rangle$. They can also be put into the machines individually or all together, as shown in Figs.~\ref{fig:CNN}(c) and (d).  In this paper, we choose $m_s=20$ for the XY model and $m_s=40$ for the XXZ model. 

The basic neural networks designed for the phase recognition task are shown in Fig.~\ref{fig:CNN}. By using Keras in the TensorFlow package\cite{tf}, all NNs begin with a convolutional layer containing 16 kernels (filters) of the size $3\times 3$ and ReLU activation functions, and then connect to four fully connected ReLU activated layers with 512, 256, 128 and 64 neurons, respectively. Finally, they are followed by an output dense layer with two (XY) or three (XXZ) neurons plus a softmax function. The output of each neuron can now be interpreted as the probability for each phase with which the input may associate. Note that the zero-padding technique on the input data is used  to keep the same input size and no pooling layers are needed due to the small size  of the input data.

In the supervised learning, the labeled training data are needed. In the XY model, we first chose two magnetic fields $h$ individually for the FM and AFM phases, each of which is expanded
within a window of size $0.1$ in the unit of $h$ at a constant anisotropy $\gamma$, to collect labeled data points. In fact, the training points can be either chosen arbitrarily deep inside each phase, or via the most confident points with the best Silhouette value in unsupervised learning approach, which will be explained in the next section. For the XXZ model, six training points are first chosen and then expand in the same manner. Here, we purposely took two points for each phase for the sake of good precisions at the inference stage. Moreover, we found that there is a more natural way to take these points by using unsupervised learning, as we will discuss later.  By setting the train-validation split ratio as 0.2, we  adopt the ADAM optimization\cite{adam} for training at learning rate $10^{-4}$ with categorical cross entropy as the loss function. Once the loss is converged after training, at the inference stage we fix whole parameters in the trained model and feed with new data for prediction. The rapid drop for the``probability'' output of the NN indicates that the trained model recognizes the phase transition from one phase to the other as a function of the input spin-spin correlation functions. 

\subsection{Supervised Results} 
\subsubsection{XY model} 

\begin{figure}[th] 
	\begin{center}
		\includegraphics[width=9cm]{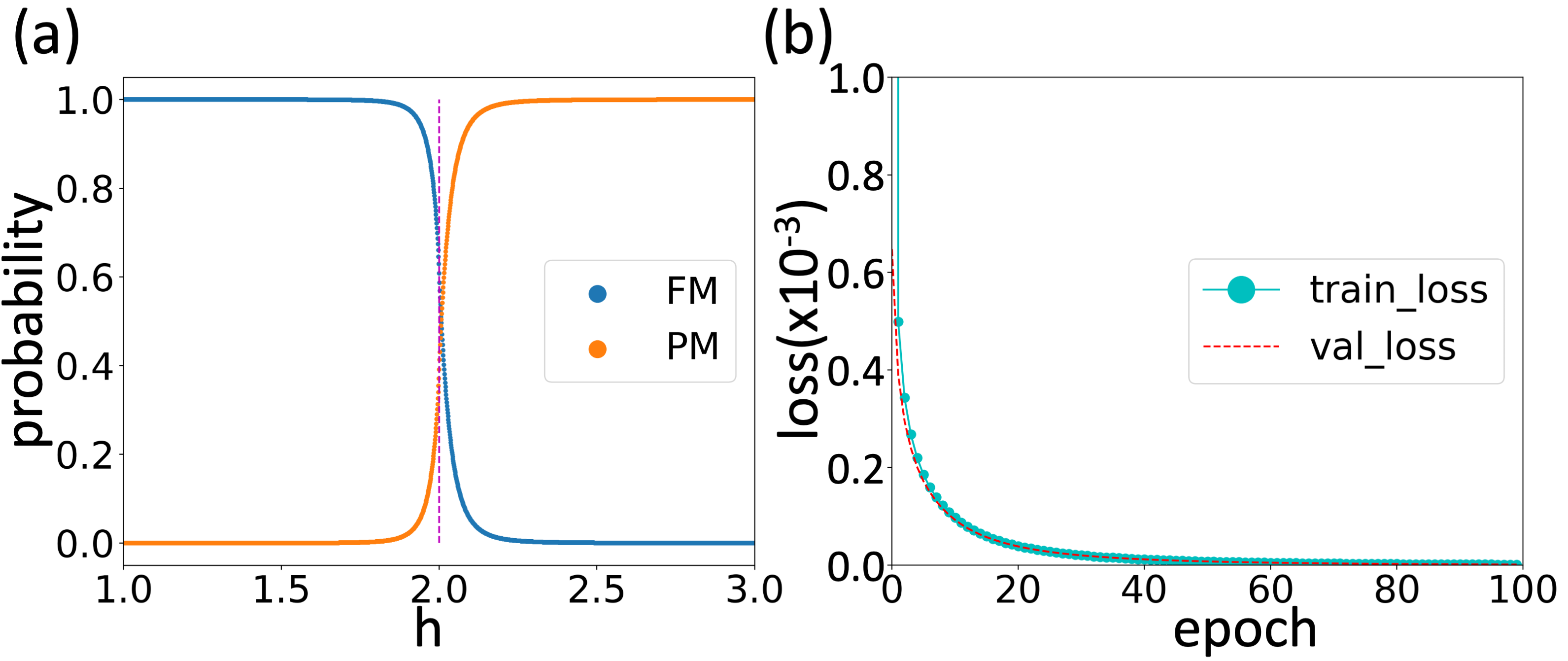}
		\caption{(a) Each neuron output of the final softmax layer corresponding to the probability of each phase, as a function of $h$ with $\gamma=0.5$ by feeding $\rho_{xx}$, $\rho_{yy}$ and $\rho_{zz}$ all together into machine. Although the training set from the correlation functions is far beyond the $h$ region shown here, the trained CNN can still recognize a quantum phase transition near $h=2$. The dashed line indicates the theoretical phase transition point. (b) The validation loss follows the trend of the training loss well, suggesting no over-fitting happened.}
		\label{fig:ResultsXY3}
	\end{center}
\end{figure}

For the supervised learning of a XY model, we choose regions around two magnetic fields $h$ as training points. For each region, a thousand of $\rho_{xx}$, $\rho_{yy}$ and $\rho_{zz}$ are calculated according to Eqs.(\ref{rhoxx}), (\ref{rhoyy}), (\ref{rhozz}) and (\ref{BA}). Those correlation functions are based on a thermodynamic system, where the system size is infinite, with a periodic boundary condition.

\begin{table}[h]
	\centering
	\begin{tabular}{|l|c|c|c|l|}\hline
		& $\rho_{xx}, \rho_{yy}, \rho_{zz}$  & $\rho_{xx}$ &
                                                                   $\rho_{yy}$    & $\rho_{zz}$   \\\hline
 	$m_s = 5$  & 1.9417    & 1.9594 & 1.9597 & 1.9032 \\
       $m_s = 10$ & 1.9510    & 1.9604 & 1.9599 & 1.9137 \\
       $m_s = 15$  & 1.9473    & 1.9599 & 1.9603 & 1.9077 \\
        $m_s = 20$  & 1.9498    & 1.9526 & 1.9528 & 1.9160 \\\hline
	\end{tabular}
	\caption{Predicted critical points by training spin-spin correlation functions of different sizes $m_s=5,10,15,20$ for the XY model. }
	\label{tab:CPsXY}
\end{table}

After training, our trained model can distinguish different phases for a given dataset and locate the phase boundaries by inputing a set of unseen data points along the magnetic field $h$. In the XY model, twenty thousand testing data points are collected uniformly from $h=1$ to $10$. By feeding $\rho_{xx}$, $\rho_{yy}$ and  $\rho_{zz}$ all together, the results are shown in Fig.~\ref{fig:ResultsXY3}(a). Typically training after 40 epochs, both training and validation losses drop to $10^{-6}$, indicating that the trained model becomes reliable, as shown in Fig.~\ref{fig:ResultsXY3}(b). The probability of the neuron output for predicting the FM phase drops from 1 to 0 at $h=2$, while the other output for predicting the PM phase arises from 0 to 1. These two curves cross each other at $h_c=1.9417, 1.9510, 1.9473, 1.9498$ for $m_s=5, 10, 15, 20$, respectively, as listed in Table~\ref{tab:CPsXY}. The predicted critical points are around $1.95$. The results for different $m_s$ are not changing too much due to the small sizes of the correlation functions. They are all within the statistical errors.

 \begin{figure}[th] 
	\begin{center}
		\includegraphics[width=9cm]{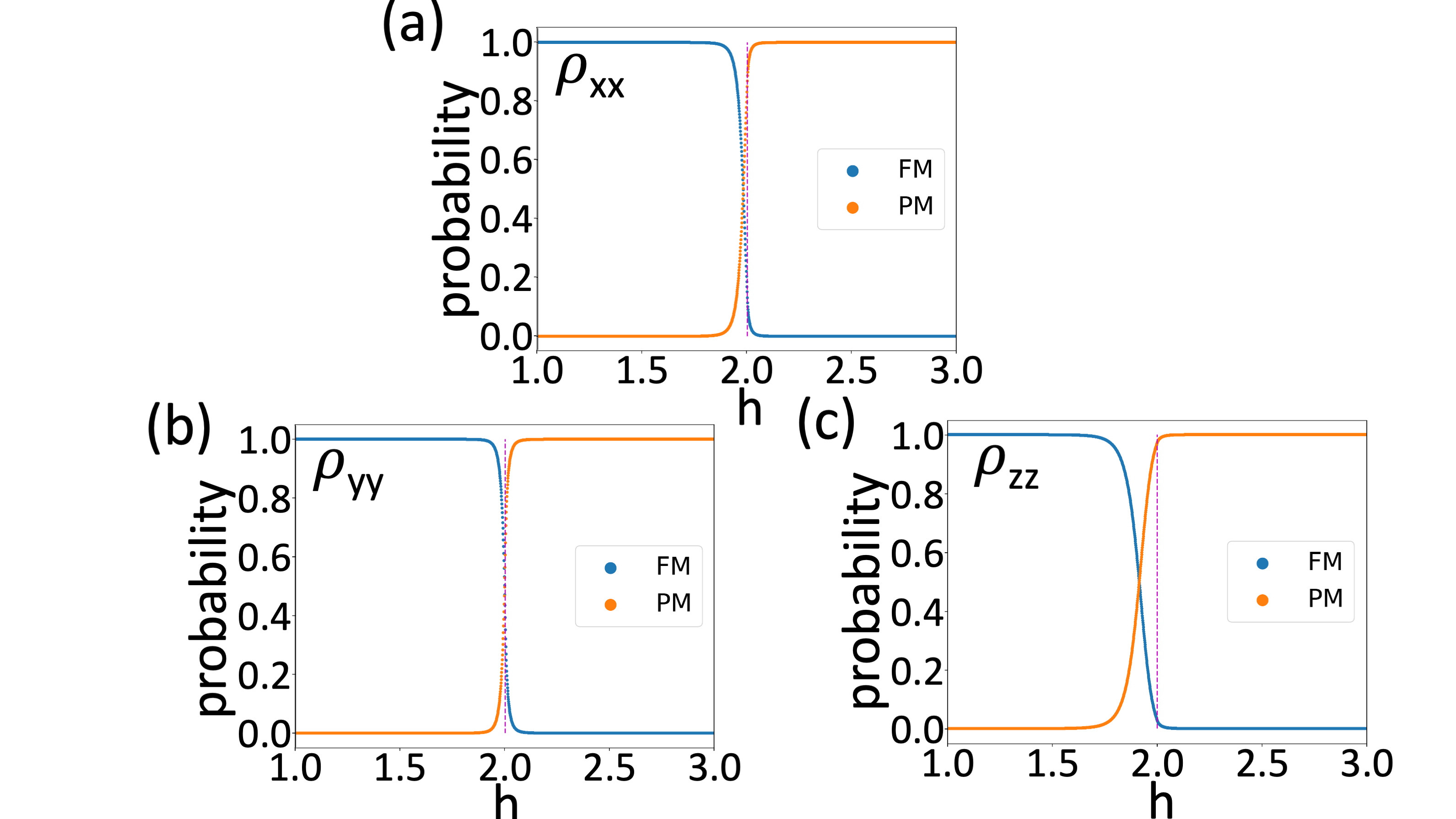}
		\caption{Results of supervised learning of the XY model by feeding (a) $\rho_{xx}$, (b) $\rho_{yy}$, or (c) $\rho_{zz}$ individually. The predictions are close to the exact value $h=2$. 
                }
		\label{fig:ResultsXY1}
	\end{center}
\end{figure}

 In Fig.~\ref{fig:ResultsXY1} we show the predicted critical points by  inputing the three correlation functions individually. The corresponding CNN architecture is shown in Fig.~{\ref{fig:CNN}}(a). In the case of $m_s=20$, $h_c$ are $1,9526, 1.9528$, and $1.9160$ with standard deviations 0.01, 0.01, and 0.02 for the training data of $\rho_{xx}$, $\rho_{yy}$ and $\rho_{zz}$, respectively. We can see that the results are almost the same as those by inputing the three  correlation functions all together. Therefore we can confirm that $\rho_{xx}$,  $\rho_{yy}$ and $\rho_{zz}$ all contain the important informations to locate the critical points.  

In Table~{\ref{tab:CPsXY}} the critical points found by inputing the correlations with difference size $m$ is shown. Essentially there exists no significant difference of the precisions among them. In other words, since DL can be viewed as a statistical model building, within the statistical errors, one can use small size of correlation functions to find the critical points with a good precision. 

\subsubsection{XXZ model}
 \begin{figure}[th] 
	\begin{center}
		\includegraphics[width=9cm]{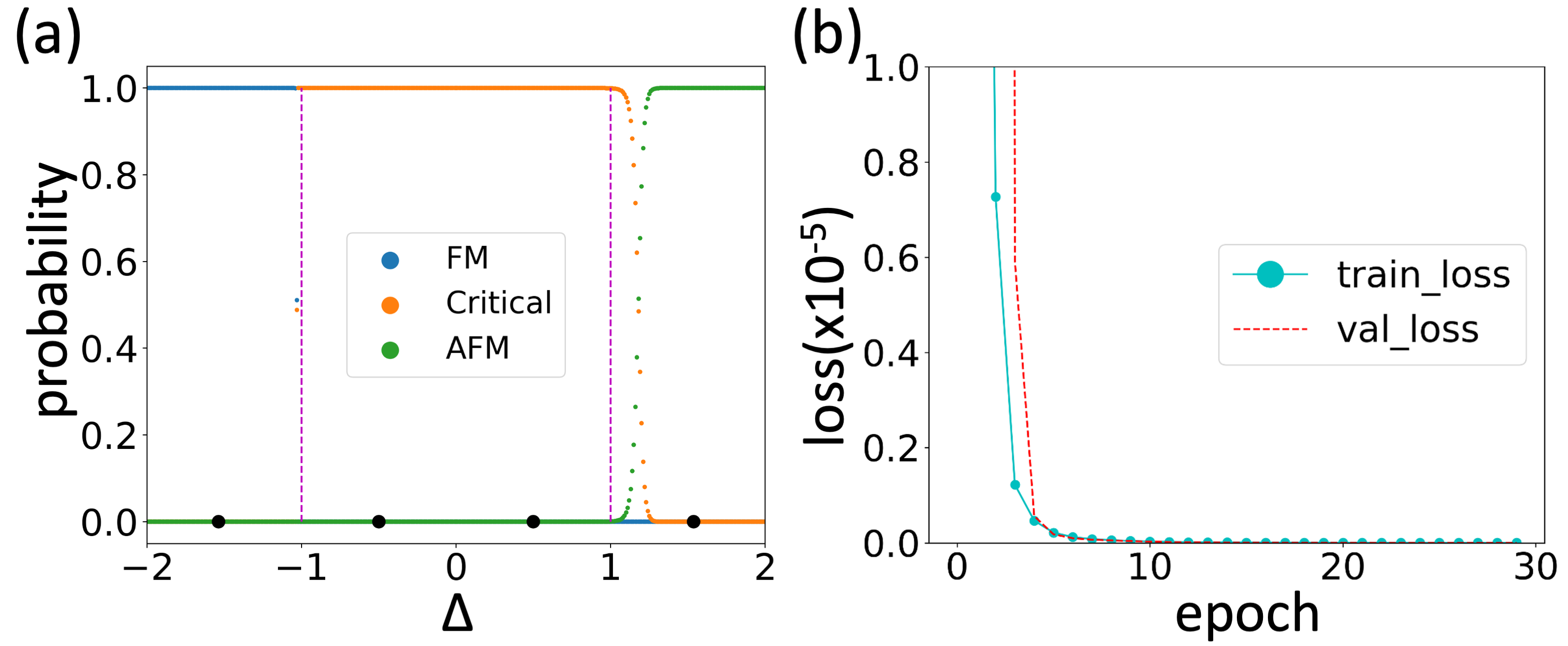}
		\caption{(a) Neuron output of the final softmax layer corresponding to the probability of each phase, as a function of $\Delta$ by feeding both $\langle S_0^{z} S_m^{z} \rangle $ and $ \langle S_0^{+} S_m^{-} \rangle$  into the trained machine. Although the training set from the correlation functions are far beyond the $\Delta$ region shown here, the CNN can still recognize quantum phase transitions near $\Delta=-1$ and $1$. The dashed lines indicate the theoretical phase transition points. (b) The validation loss follows the trend of the training loss well, suggesting no over-fitting happened.}
		\label{fig:ResultsXXZ2}
	\end{center}
\end{figure}

For the supervised learning of the XXZ model, we take two anisotropies, {\it i.e.} $\Delta$ values, for each phase as training point centers, and thus there are six in total. Around each $\Delta$, one thousand of $\langle S_0^{z} S_m^{z} \rangle $ and  $ \langle S_0^{+}S_m^{-} \rangle$ are calculated by MPS Toolkit of iDMRG\cite{Ian2}. 

Same as that for the XY model, after training, the machine can serve as a well-trained model to find out the TPs by feeding a set of unseen data points along $\Delta$. Here, we also take two thousand testing data points for exploring phase transitions. By inputing both  $\langle S_0^{z} S_m^{z} \rangle $ and $ \langle S_0^{+} S_m^{-} \rangle$, we show the predicted critical points in Fig.~{\ref{fig:ResultsXXZ2}}(a).  Typically after 14 epochs of training, the training and validation losses decrease to less  than $10^{-8}$ as shown in Fig.~{\ref{fig:ResultsXXZ2}}(b). Namely, the trained model has found the optima for finding those critical points. The probability of predicting the FM phase drops from 1 to 0 around the critical point $\Delta_c = -1$, whereas that of predicting the critical phase raises from 0 to 1. Similar situation also happens for the critical point between the critical and AFM phases. The predicted critical points $(\Delta_c^1, \Delta_c^2)$, which are taken at the crossing points of the probabilities, are $(-1.0542, 1.1922), (-1.0303, 1.1904), (-1.0471, 1.1961)$ and $(-1.0323, 1.2078)$ for $m_s=10, 20, 30$
and $m_s=40$, respectively, as shown in Table.~\ref{tab:CPsXXZ}. The predicted $\Delta_c^1$ are closer to $-1$, whereas the differences between the predicted $\Delta_c^2$ and the real critical point $1$ are bigger. In order to understand the situation, we feed  $\langle S_0^{z} S_m^{z} \rangle $ and $ \langle S_0^{+} S_m^{-} \rangle$ separately into machines. 

\begin{table}[h]
	\centering
	\begin{tabular}{|l|c|c|c|c|c|l|}\hline
		& \multicolumn{2}{c|}{$\langle S_0^{z} S_m^{z} \rangle $,
                  $\langle S_0^{+} S_m^{-}\rangle$}  &
                   \multicolumn{2}{c|}{$\langle S_0^{z}  S_m^{z} \rangle$} &
                                                                    \multicolumn{2}{c|}{$\langle
                                                                            S_0^{+}
                                                                    S_m^{-}
                                                                    \rangle$}
          \\\hline
         & $\Delta_c^{1}$& $\Delta_c^{2}$ &  $\Delta_c^{1}$&
                                                             $\Delta_c^{2}$
                         &  $\Delta_c^{1}$& $\Delta_c^{2}$ \\\hline
 	$m_s = 10$   & -1.0542       & 1.1922 & -1.0894     & 1.1734     & -1.0604     & 1.1984 \\
    $m_s = 20$ & -1.0303       & 1.1904  & -1.0168     & 1.2419     & -1.0348     & 1.1479 \\
    $m_s= 30$   & -1.0471       & 1.1961  & -1.0192     & 1.2729    & -1.0550     & 1.1214  \\
    $m_s = 40$  & -1.0323       & 1.2078 & -1.0182     & 1.2925 & -1.0492     & 1.0991 \\\hline
	\end{tabular}
	\caption{Predicted critical points by training spin-spin correlation functions of different sizes $m_s=10, 20, 30, 40$ for the XXZ model. }
	\label{tab:CPsXXZ}
\end{table}

\begin{figure}[th] 
	\begin{center}
		\includegraphics[width=9cm]{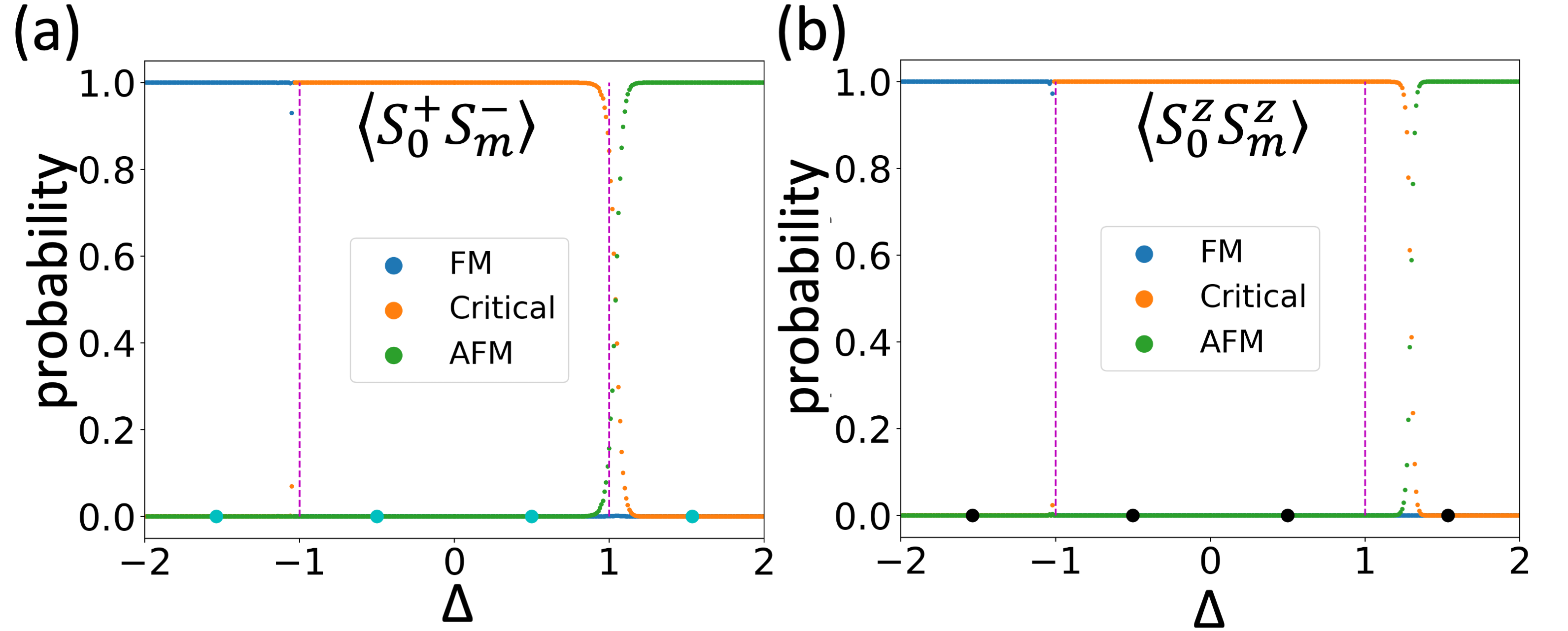}
		\caption{Results of supervised learning of the XXZ model by feeding (a) $\langle S_0^{+} S_m^{-}\rangle $ and (b) $ \langle S_0^{z} S_m^{z} \rangle$, separately, with $m_s=40$. Please refer to Section III C for more discussions.}
		\label{fig:ResultsXXZ1}
	\end{center}
\end{figure}

Fig.~\ref{fig:ResultsXXZ1} shows the probability of predicting the critical points by feeding the correlation functions  $\langle S_0^{z}S_m^{z} \rangle $ and $ \langle  S_0^{+} S_m^{-} \rangle$ separately. One can also see the predicted values in Table~\ref{tab:CPsXXZ} for different subsystem size $m_s=10,20,30, 40$. 
For the first order phase transition from the FM phase to the critical one, {\it i.e.} $\Delta=-1$,
either $\langle S_0^{z} S_m^{z} \rangle $ or $ \langle  S_0^{+}S_m^{-} \rangle$ can predict more precise TPs, however, for the second order TPs $\Delta_c =1$, $ \langle  S_0^{+}S_m^{-} \rangle$ is getting more and more precise through the increasing $m_s$, whereas the predicted values by feeding $\langle S_0^{z} S_m^{z} \rangle $ are not so precise throughout different
subsystem sizes.  

The aforementioned phenomena can be understood as follows. Near any first order TPs, the gaps are opened faster, therefore the machine can easily recognize TPs, while near the second order TPs, the gaps are opened much more slowly, and thus the precision for finding such TPs is poorer from both correlations. However, when going into the AFM phase,  $\langle S_0^{z} S_m^{z} \rangle $ shows a smaller growth than the decrease of $ \langle  S_0^{+} S_m^{-} \rangle$ and it results in bigger prediction errors for $\Delta_c^2$. These results give us a hint, why the predicted critical points $\Delta_c^2$ by inputing two correlation functions together are in between of the predicted values by feeding each correlation individually. The more imprecise results obtained by $\langle S_0^{z} S_m^{z} \rangle $ weaken the prediction precision when two correlation functions are input spontaneously.  

The results for the XXZ model are not so precise than those for the XY model due to the fact that we use numerics (iDMRG) to calculate the correlations for the XXZ model, instead of using analytical expressions as in the XY model. However, the results are still encouraging. Especially the predicted critical points obtained from $ \langle S_0^{+} S_m^{-} \rangle$ are surprisingly in good accordance with the analytical results when $m_s\le 20$. The fact shows the advantage and the effectiveness of our approach for searching the patterns in the correlation functions. 

\section{Unsupervised Learning}

\subsection{ML Algorithms}
One of the ultimate goals in ML is to discover the hidden patterns behind a given data without any human intervention or manual labeling. This kind of ML algorithms are called unsupervised learning. Since one does not need any label tagging in the process of unsupervised learning, it is usually difficult and commonly considered as a kind of holy grail in the science community. In our previous study\cite{ChungTsai1DUSL}, we already demonstrated a working procedure to obtain the TPs by mainly unsupervised learning, and thus we just introduce our method here in a simpler version.

\begin{figure}[th] 
	\begin{center}
		\includegraphics[width=9cm]{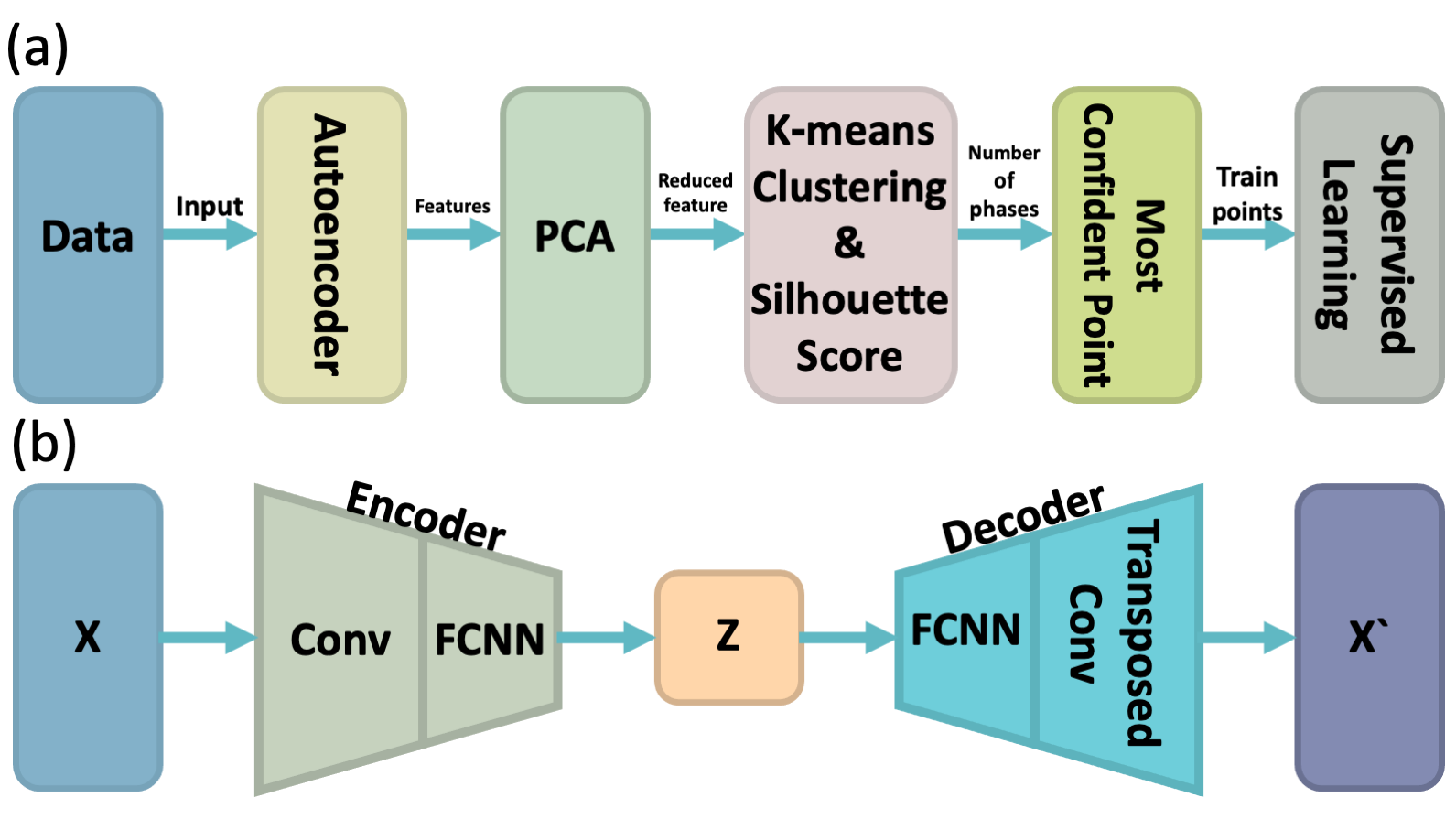}
		\caption{(a) The proposed working protocol to identify different phases and to finely determine the phase boundaries without prior knowledge. (b) The schematic model architecture of the autoencoder (AE). (Conv: Convolution module; FCNN: Fully connected neural network module.) }
		\label{fig:ModelArchUSL}
	\end{center}
\end{figure}

The proposed working protocol is schematically shown in Fig.~\ref{fig:ModelArchUSL}(a). There are four steps: (1) The input correlation functions are fed into an autoencoder (AE) for unsupervised training in order to extract effective features via dimension reduction. (2) The necessary feature dimension is determined by principal component analysis (PCA) when $99\%$ of total variance of input features is kept. (3) The total number of phases is then determined by K-means clustering for the extracted, necessary features, followed by silhouette analysis (SA). (4) Sharp phase boundaries can be further obtained with the help of supervised learning that we constructed in the last section. We explain a few key algorithms below.

An autoencoder (AE) compresses input data into more efficient representation in an unsupervised way. It consists of two parts, an encoder and a decoder. A typical model architecture of AE is shown in Fig.~\ref{fig:ModelArchUSL}(b). It is made of a convolution layer followed by a linear module composed of fully connected hidden layers. In the decoder, it is arranged in a reversed manner to that of the encoder, except that now the convolution layer is replaced by a transposed convolutional one. 

As opposed to AE, principal-component analysis (PCA) is a linear method to reduce the dimension and to visualize the data. In order to do that, PCA uses an orthogonal and linear transformation of the input features to a sorted set of new variables by their variance. 

Once obtained the extracted features from the input data, $K$-means algorithm is a clustering algorithm without any supervision. Given the number of clusters $n$, K-means is used to find out the best cluster formation such that the variance within each cluster is minimized. 

However, in order to use K-means to find out the best cluster formation, one still has to provide the number of clusters $n$. To find out $n$ automatically, we employ silhouette analysis (SA). For a give set of clusters, SA assigns a value, called the silhouette value $s(x)$, which is bounded between $+1$ and $-1$, to each data point within a cluster. $s(x)$ can be interpreted as a measure of how alike $x$ is to its own cluster (cohesion) compared to the other one (separation). One computes $s$-score, namely, the mean of $s(x)$, as a function of the number of clusters $n$ after K-means clustering, and then we take the best $n$ as the one giving the maximum $s$-score. 

Once the best choice for $n$ is given, the data point with the highest silhouette value within the same cluster can serve as the most confident point to build a ``labeled'' training set to train a
supervised neural network. It can improve the precision of the predicted critical points originally obtained in an unsupervised learning over the parameter space. 

\subsection{Unsupervised Results}
\subsubsection{XY model}

\begin{figure}[th] 
	\begin{center}
		\includegraphics[width=9cm]{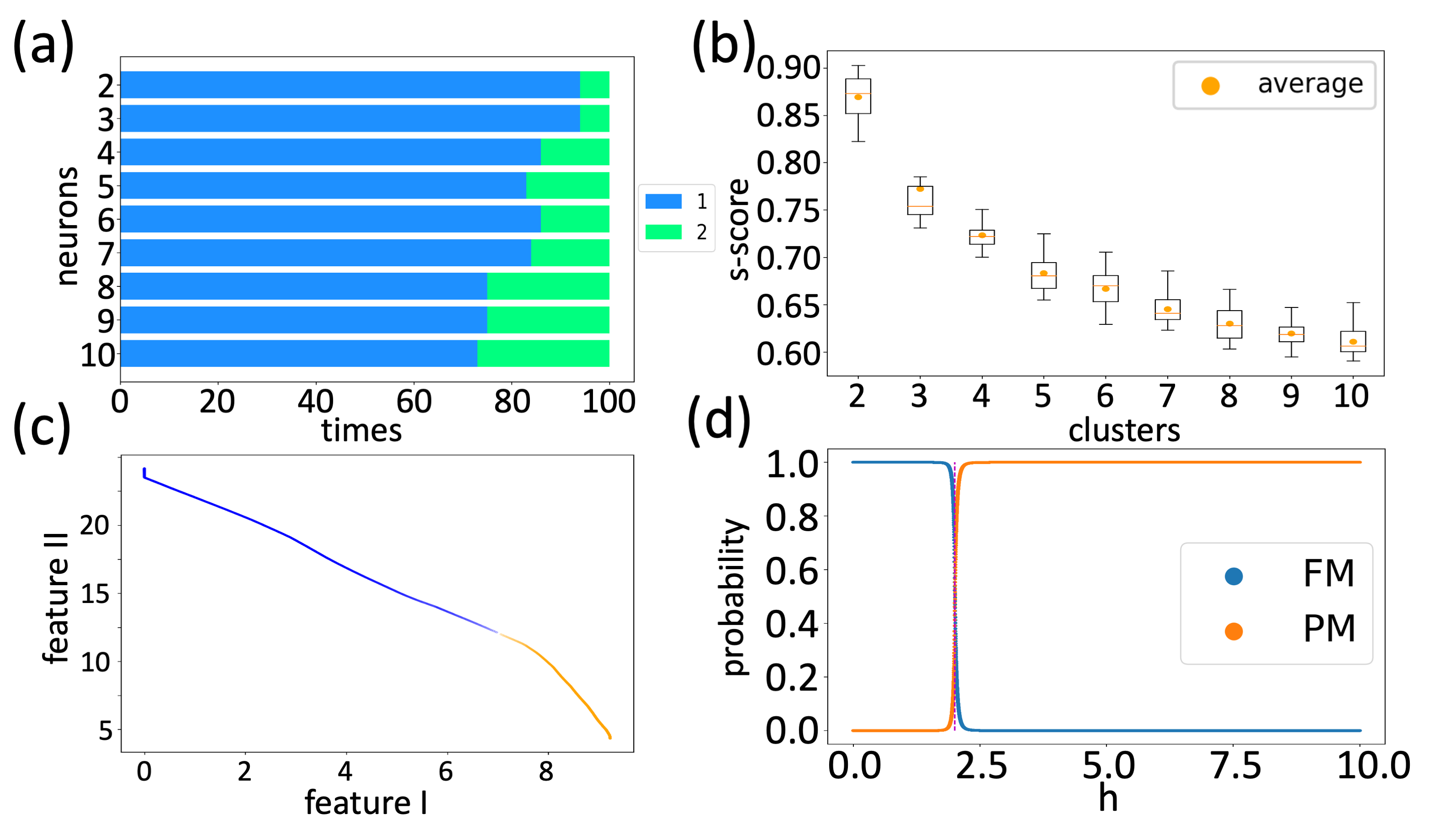}
		\caption{AE results for the combined spin-spin correlation functions ($\rho_{xx}$, $\rho_{yy}$, $\rho_{zz}$) of the XY model as the input data. (a) The discrete distribution of the necessary number of neurons $d_z$ for a given $n_{mid}$ of neurons in the middlemost layer (2 to 10 along $y$ axis). Results from 100 independently trained AE are statistically calculated: The length of every color bar is proportional to the number of times that $d_z$ occurred within 100 models. Different colors in the legend represent different $d_z$. (b) The box plot of the $s$-scores as a function of $n$ clustering (via K-means method). (c) Latent representations projected to a subspace spanned by the first two principal components (feature map). Each color indicates its corresponding cluster (phase). (d) The neuron output (phase diagram) as a function of $h$ with $m_s=20$, $\gamma=0.5$ for 1D XY model from a trained CNN by supervised learning in the last step of the ML protocol.  The dashed line indicates the theoretical phase transition boundary.}
		\label{fig:USResultsXY3}
	\end{center}
\end{figure}

\begin{figure}[th] 
	\begin{center}
		\includegraphics[width=9cm]{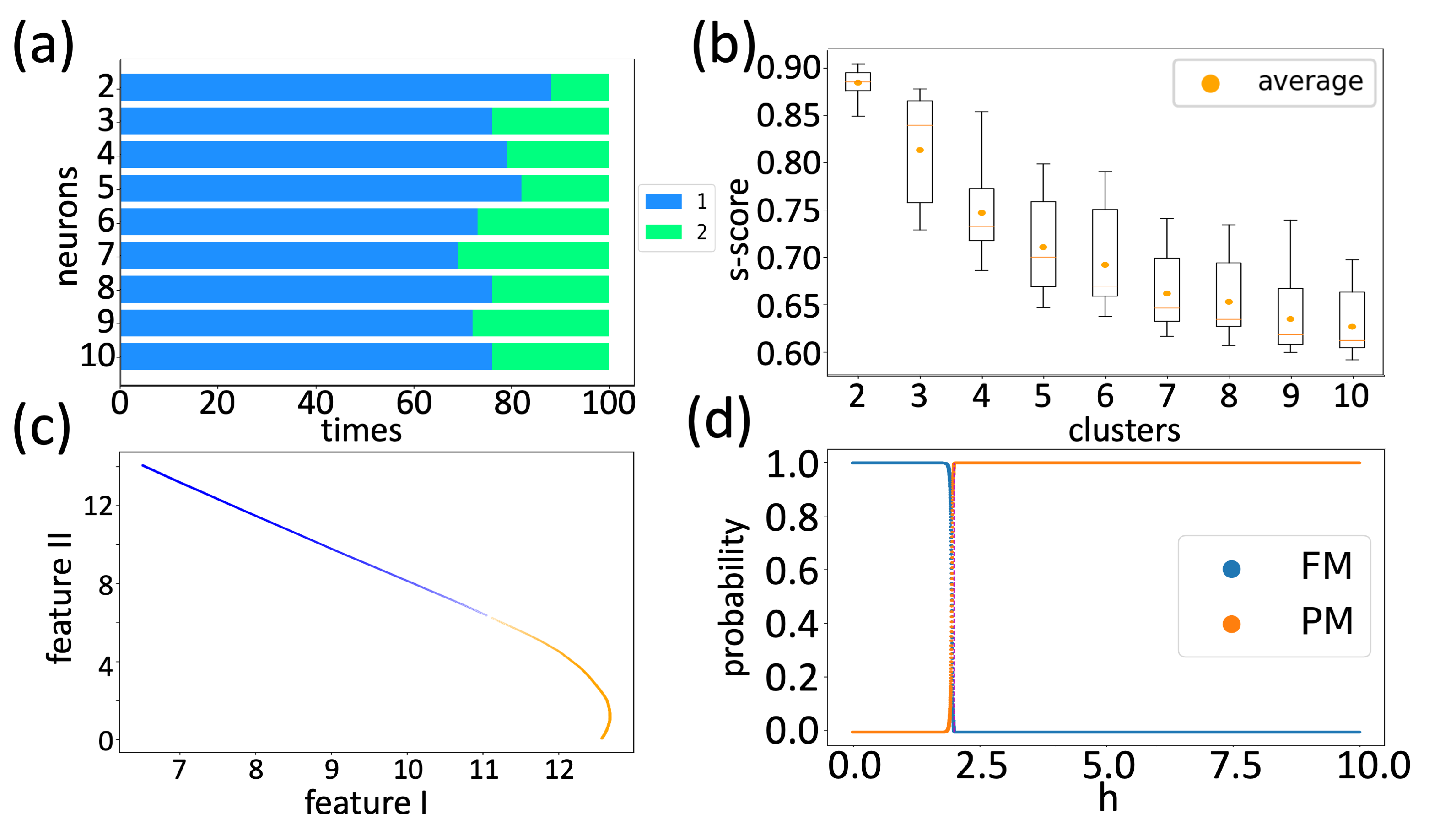}
		\caption{Same as Fig.~\ref{fig:USResultsXY3}, except that here, $\rho_{xx}$ function is taken as the
			input data. }
		\label{fig:USResultsXYxx}
	\end{center}
\end{figure}

\begin{figure}[th] 
	\begin{center}
		\includegraphics[width=9cm]{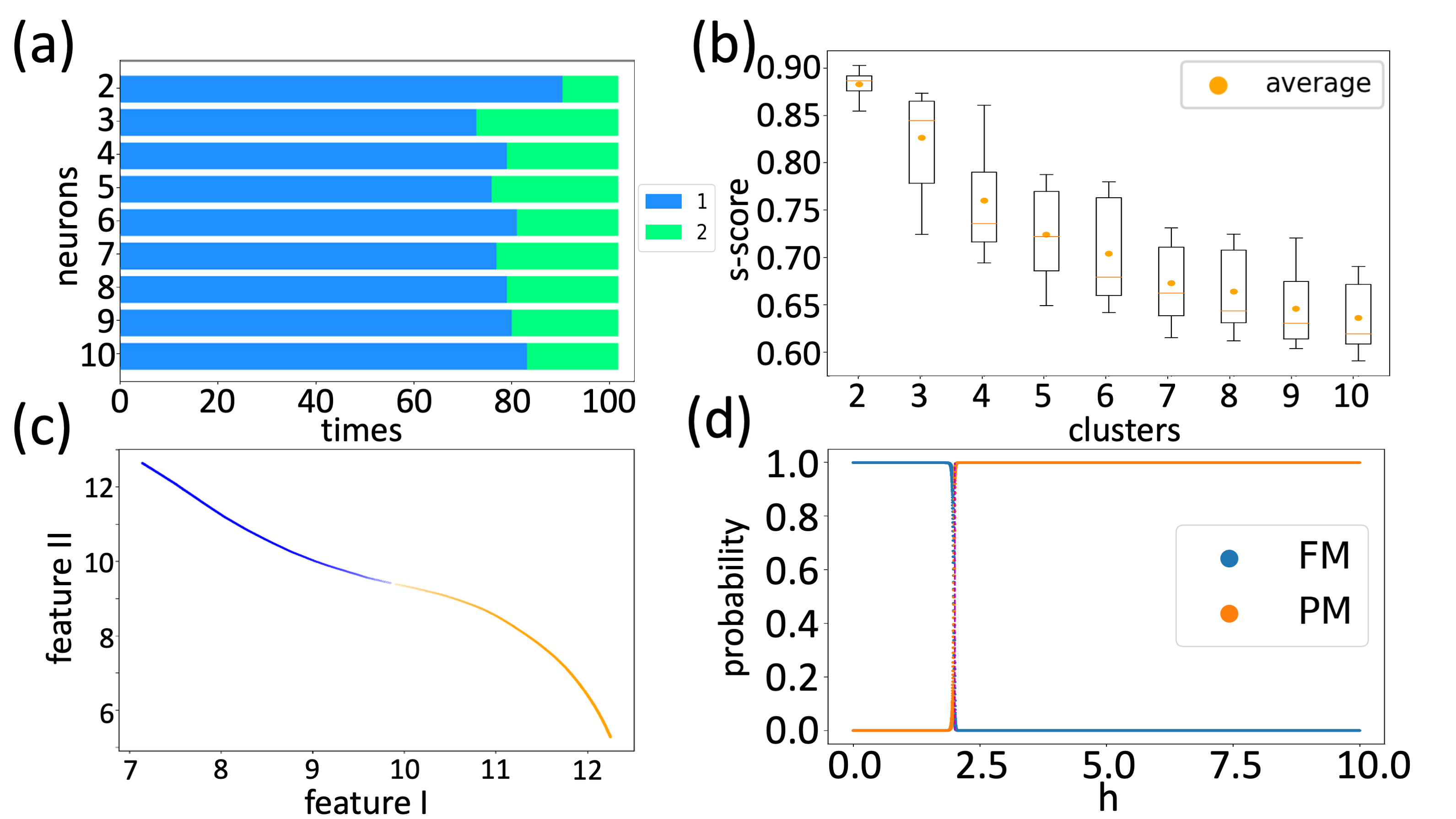}
		\caption{ Same as Fig.~\ref{fig:USResultsXY3}, except that here, $\rho_{yy}$ function is taken as the
				input data. }
		\label{fig:USResultsXYyy}
	\end{center}
\end{figure}

\begin{figure}[th] 
	\begin{center}
		\includegraphics[width=9cm]{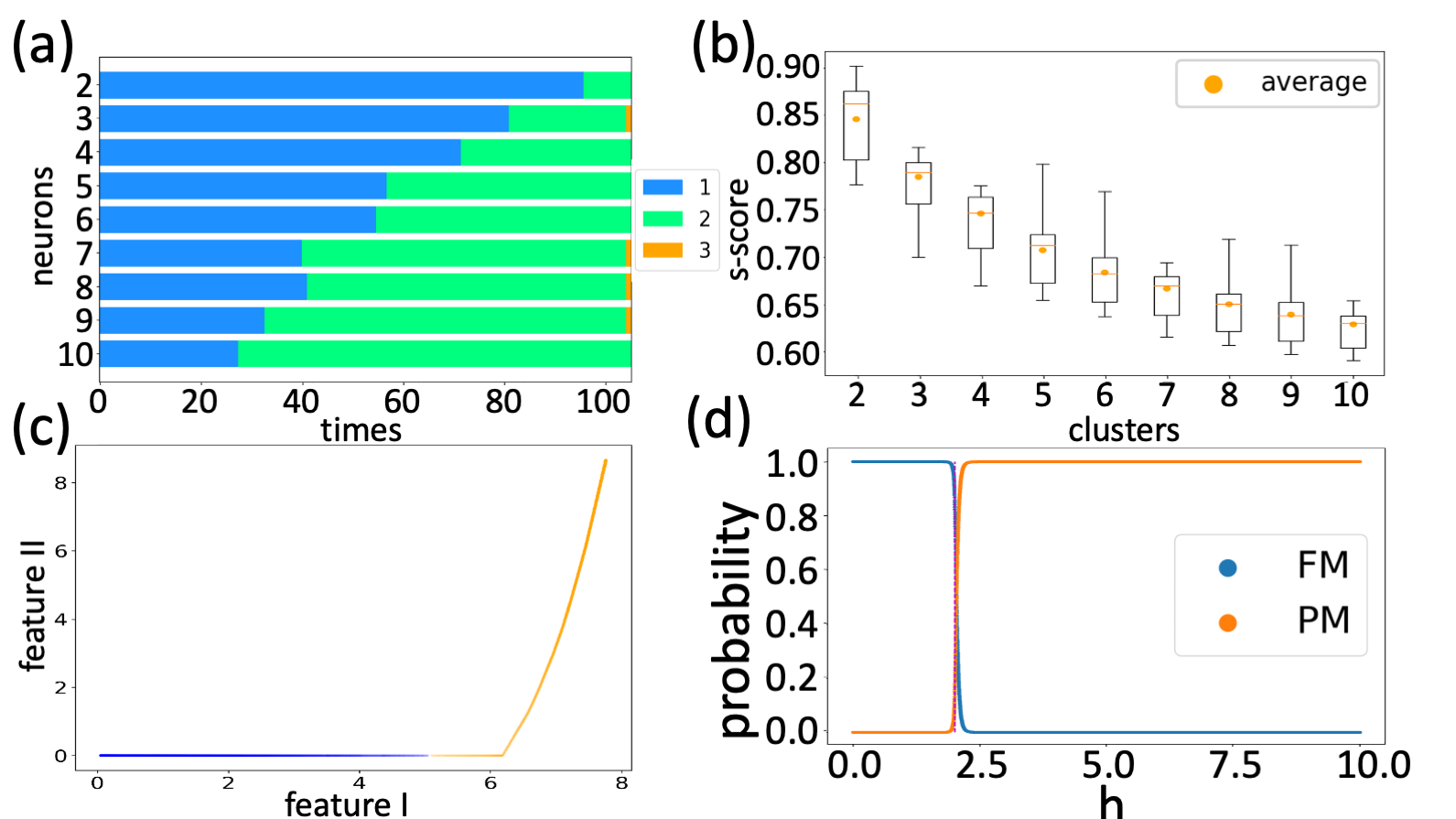}
		\caption{ Same as Fig.~\ref{fig:USResultsXY3}, except that here, $\rho_{zz}$ function is taken as the
				input data. }
		\label{fig:USResultsXYzz}
	\end{center}
\end{figure}

We first prepare the input image dataset of correlation functions by generating 10001 $\rho_{xx}, \rho_{yy}$ and $\rho_{zz}$ with the same method described in the supervised learning at evenly divided magnetic field $h$ from $0$ to $10$, with subsystem size $m_s=20$ of an infinite
chain with periodic boundary conditions. We feed the three correlation functions all together or separately. In Fig.~\ref{fig:USResultsXY3} we show the results by feeding three correlation functions at the same time, while Figs.~\ref{fig:USResultsXYxx},  \ref{fig:USResultsXYyy} and \ref{fig:USResultsXYzz} show the results by inputing $\rho_{xx}$, $\rho_{yy}$ and $\rho_{zz}$, respectively, into the machine. 

Following the ML protocol described in ML algorithms, we firstly have to train a neural network to encode our input data to effective representations in the latent space. In order to discover the minimum dimension $d_z$ of the latent space, a series of AEs with the same model architecture is trained except for the number of hidden neurons in the middlemost layer $n_{mid}$ from $2$ to $10$. For each $n_{mid}$, the necessary dimension of the converged latent representation to keep at least $99\%$ variance of them by PCA is recorded. For doing the statistical analysis, we repeat $100$ times the same training procedures with the same initial weight distribution. In this way, the minimun dimension $d_z$ is decided if such number becomes indispensable (dominant) in the discrete distribution of $d_z$ when $n_{mid}$ increases.  
   
In order to obtain the best number of clusters $n$, we utilize SA through the K-means algorithm. In the following calculations, we already fixed $n_{mid} = d_z=2$ (fed data: three correlations, $\rho_{xx}$, $\rho_{yy}$) or $n_{mid} =d_z=3$ ($\rho_{zz}$)  hidden neurons in the middlemost layers as suggested by (a) of Figs.~\ref{fig:USResultsXY3}, \ref{fig:USResultsXYxx}, \ref{fig:USResultsXYyy}, and \ref{fig:USResultsXYzz}, respectively. In (b) of those four figures, the mean silhouette values achieve the higgest one when $n=2$, suggesting that there are two different phases (clusters) for the XY model.

In (c) of Figs.~\ref{fig:USResultsXY3}, \ref{fig:USResultsXYxx}, \ref{fig:USResultsXYyy}, and \ref{fig:USResultsXYzz}, we show the projection of latent representations into a 2D space spanned by the first two principal components (features). These plots show us how the system could be separated into two clusters (phases). The feature plot is a continuous curve, which characterizes a second order phase transition. We will see the differences of such feature plots between the first and the second order phase transition in the next subsection. 

By going through the aforementioned recipe, the TPs are statistically found at the mean values 2.0314, 1.9742, 1.9752, 2.0867 with the standard deviations 0.03, 0.02, 0.02, and 0.14 for Figs.~\ref{fig:USResultsXY3}, \ref{fig:USResultsXYxx},  \ref{fig:USResultsXYyy}, and \ref{fig:USResultsXYzz}, respectively, after collecting clustering results from 100 sets of latent representations via different trained AEs. 

\begin{figure}[th] 
	\begin{center}
		\includegraphics[width=7cm]{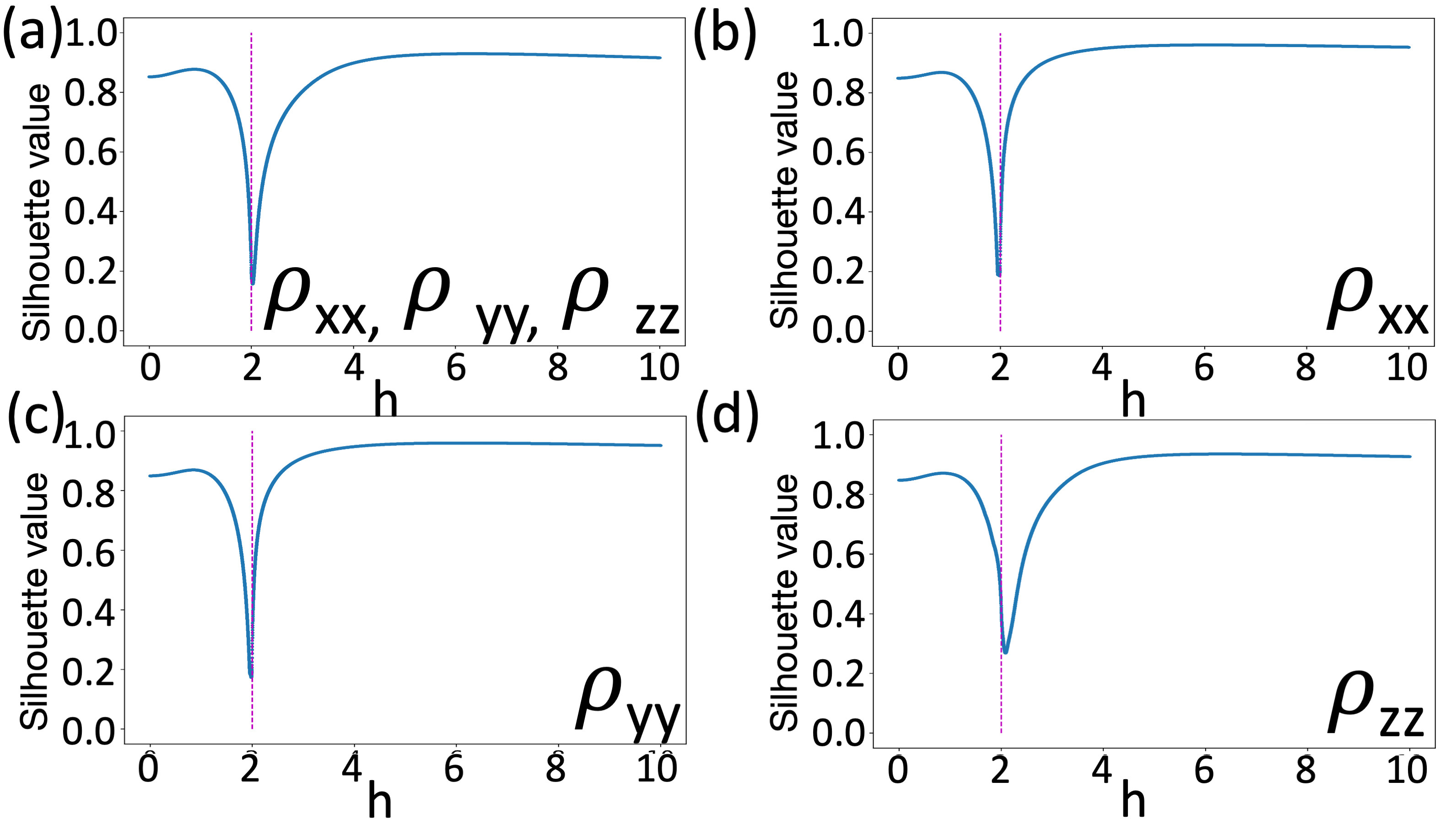}
		\caption{ Silhouette values as a function of $h$ for $\gamma=0.5$ and $m=20$ with (a) $\rho_{xx}, \rho_{yy}$ and $\rho_{zz} $ all together, (b) $\rho_{xx}$, (c) $\rho_{yy}$, and (d) $\rho_{zz}$, as the input data. Notice that the minima of Silhouette values are close to the phase transition point $h=2$. }
		\label{fig:SAXY}
	\end{center}
\end{figure}

The silhouette values of the XY model are drawn in Fig.~\ref{fig:SAXY}. One can see that the minima of silhouette values almost lie on the TPs. This suggests that SA can serve as a good indicator for phase transitions. In order to find more accurate TPs, we first take two representative points, which obtain the highest s-score in each cluster. We call them the most confident points (MCPs). In our case, they are (0.89, 6.31), (0.85, 6.15), (0.85, 6.13) and (0.87, 6.34) for the input data as the combined three correlations, $\rho_{xx}$, $\rho_{yy}$, and $\rho_{zz}$, respectively. For each phase, we expand symmetrically around the MCP by a window of $0.1$ to get 1000 points with equal spacing. These 2000 data points form our training dataset with labels, and the original 10,001 ones become our test dataset (without labels). By further training CNN models with these labeled datasets, the PTs found becomes sharper at mean values, 1.9498, 1.9526, 1.9528 and 1.9160 with smaller standard deviation, 0.0272, 0.0129, 0.0128 and 0,0215, as shown in Figs.~\ref{fig:USResultsXY3}, \ref{fig:USResultsXYxx}, \ref{fig:USResultsXYyy}, and \ref{fig:USResultsXYzz}, respectively, at the testing stage.

\subsubsection{XXZ model} 

\begin{figure}[th] 
	\begin{center}
		\includegraphics[width=9cm]{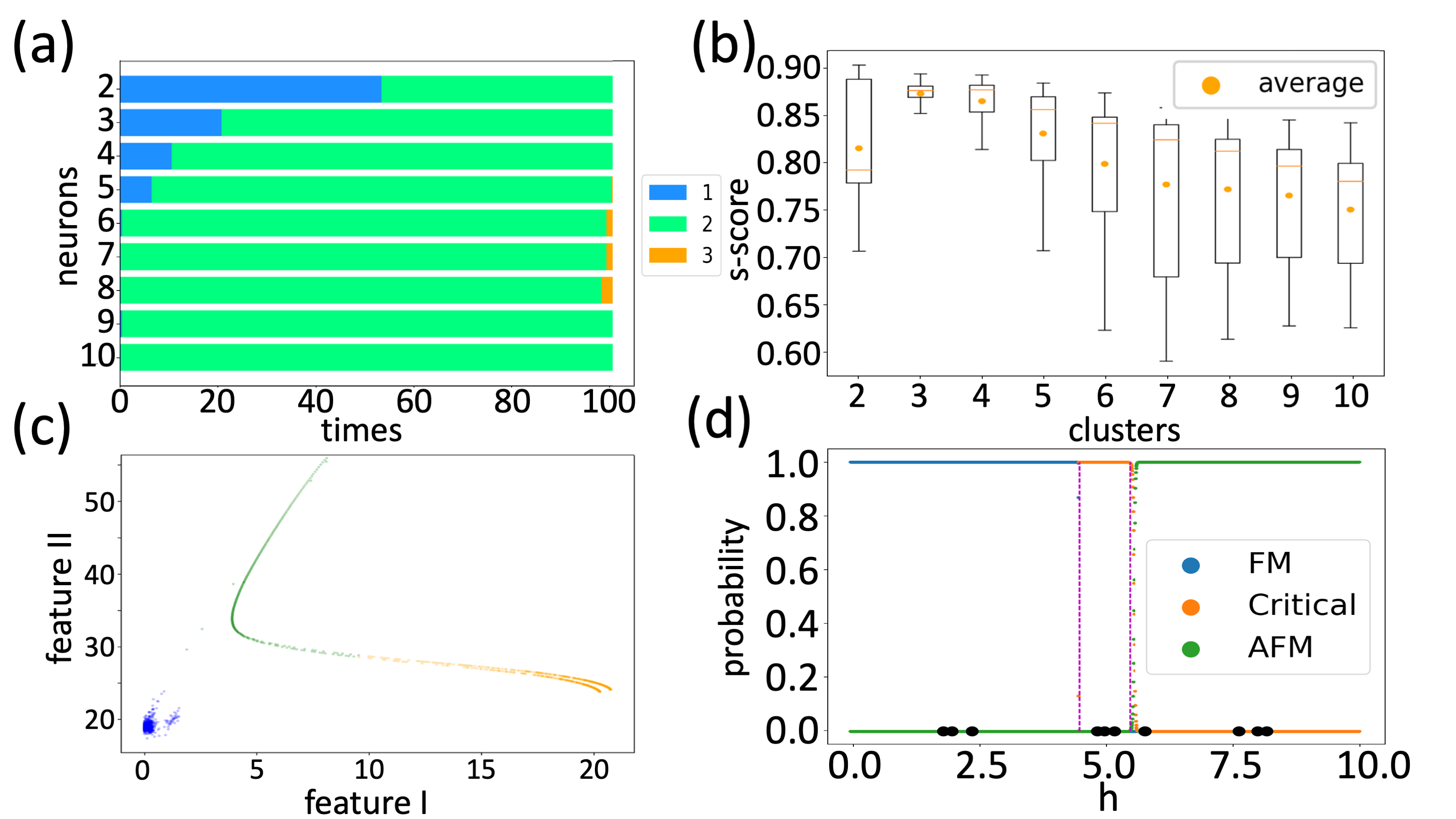}
		\caption{ AE results for both spin-spin correlation functions ($\langle S_0^{+} S_m^{-} \rangle$ and  
			$\langle S_0^{z} S_m^{z} \rangle$) of the XXZ model as the input data. (a) The discrete distribution of the necessary number of neurons $d_z$ for a given $n_{mid}$ of neurons in the middlemost layer (2 to 10 along $y$ axis). Results from 100 independently trained AE with both correlation functions are statistically calculated: The length of every color bar is proportional to the number of times that $d_z$ occurred within 100 models. Different colors in the legend represent different $d_z$. (b) The box plot of the $s$-scores as a function of $n$ clustering (via K-means method). (c) Latent representations projected to a subspace spanned by the first two principal components (feature map). Each color indicates its corresponding cluster (phase). It is obvious that for the first-order phase transition, the two clusters are disconnected, whereas for the second-order phase transition, they are continuously connected. (d) The neuron output (phase diagram) as a function of $\Delta$ for 1D XXZ model from a trained CNN by supervised learning in the last step of the ML protocol. We choose ten seeds (gray dots) as our labeled training data. The dashed lines indicate the theoretical phase transition boundaries. Please see the main text and discussions for more details.}
		\label{fig:USResultsXXZ2}
	\end{center}
\end{figure}

\begin{figure}[th] 
	\begin{center}
		\includegraphics[width=9cm]{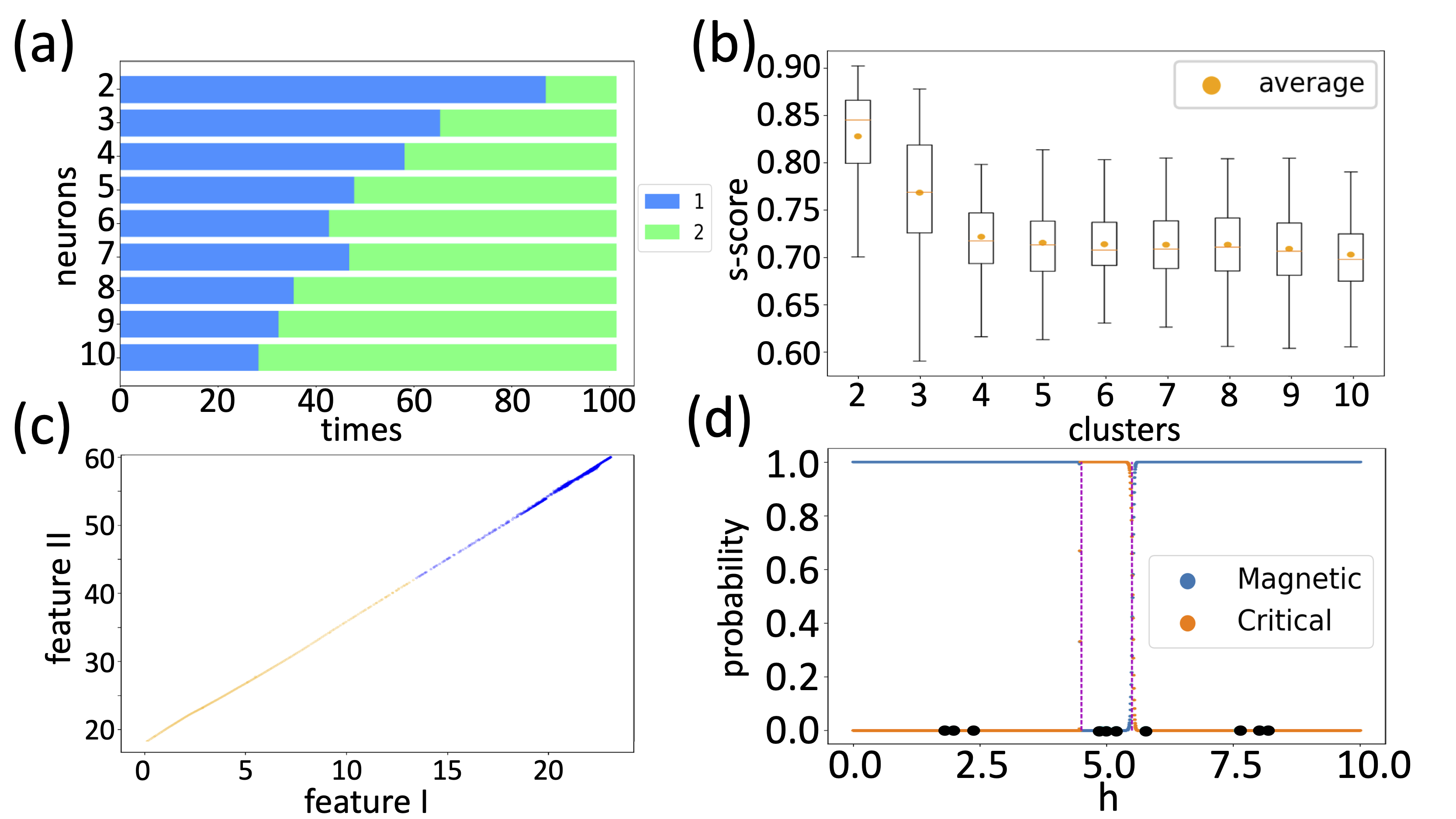}
		\caption{Same as Fig.~\ref{fig:USResultsXXZ2}, except that here, $\langle S_0^{+} S_m^{-} \rangle$ is taken as the input data. Moreover, in (b), since $\langle S_0^{+} S_m^{-} \rangle$ is not relevant in the FM and AFM phases, {\it i.e.} they can not distinguish FM from AFM, they show only two clusters (critical and magnetic phases); in (c), unlike the input data with $\langle S_0^{z} S_m^{z} \rangle$, there exists no signature of the first-order phase transition if we only feed $\langle S_0^{+} S_m^{-} \rangle$ into the  machine, therefore the two clusters are continuously connected. Please see the main text and discussions for more details. }
		\label{fig:USResultsXXZpm}
	\end{center}
\end{figure}

\begin{figure}[th] 
	\begin{center}
		\includegraphics[width=9cm]{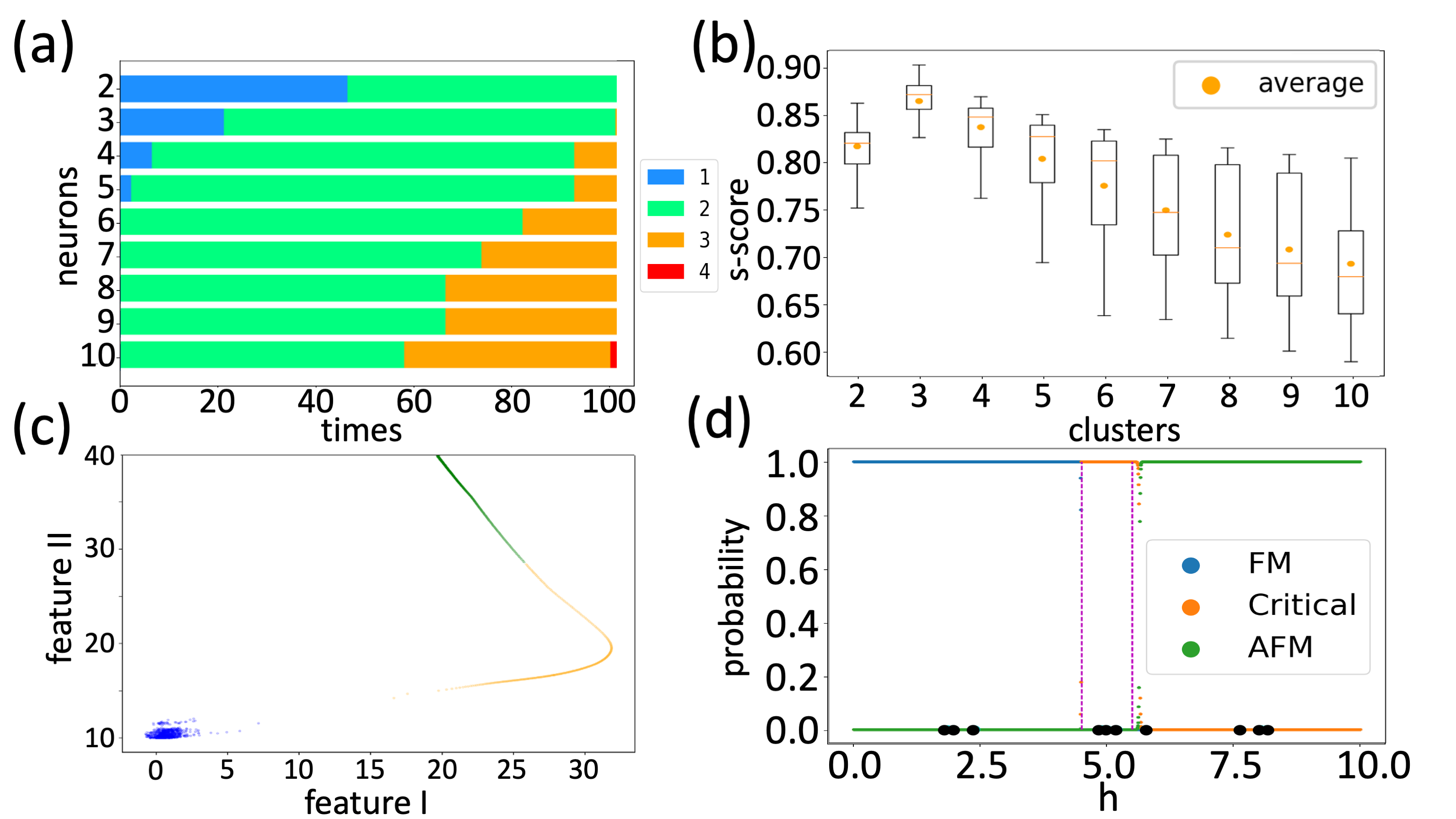}
		\caption{Same as Fig.~\ref{fig:USResultsXXZ2}, except that here, $\langle S_0^{z} S_m^{z} \rangle$ is taken as the input data. Moreover, in (b), it is obvious that for the first-order phase transition, the two clusters are disconnected, whereas for the second-order phase transition, they are continuously connected. Please see the main text and discussions for more details. }
		\label{fig:USResultsXXZzz}
	\end{center}
\end{figure}

For the XXZ model, we  prepare the input ``images'' by generating 20001 $\langle S_0^{+} S_m^{-}\rangle$ and $\langle S_0^{z} S_m^{z}\rangle$ by iDMRG at evenly divided  $\Delta$ from $-10$ to $10$. In unsupervised learning we choose the subsystem size $m_s=40$. Similar to the case of XY model, we input both correlation functions $\langle S_0^{+} S_m^{-} \rangle$ and  $\langle S_0^{z} S_m^{z} \rangle$
together or separately. In Fig.~\ref{fig:USResultsXXZ2} we show the results by feeding both correlation functions, whereas in Figs.~\ref{fig:USResultsXXZpm} and \ref{fig:USResultsXXZzz} the results by inputing $\langle S_0^{+} S_m^{-} \rangle$ and  $\langle S_0^{z} S_m^{z}\rangle$, respectively, into machines are shown. 

Similar to the XY model, we first train a neural network to encode our data of correlation functions to the latent representation. We determine the minimal dimension $d_z$ of the latent space as follows. Firstly, we train 100 AEs with the same model architecture for each given number of hidden neurons in the middlemost layer ($n_{mid}$, from 2 to 10). In (a) of Figs.~\ref{fig:USResultsXXZ2}, \ref{fig:USResultsXXZpm} and \ref{fig:USResultsXXZzz}, we then compute the discrete distribution of $d_z$, {\it i.e.}, the necessary dimensions of the latent representations to keep at least $99\%$ variance of all encoded representation vectors for each $n_{mid}$ by PCA. $d_z$ is finally suggested to be $3$ for the input data using both
correlation functions in Fig.~\ref{fig:USResultsXXZ2}, 2 for those using $\langle S_0^{+} S_m^{-} \rangle$  in Fig.~\ref{fig:USResultsXXZpm}, and 4 for those using $\langle S_0^{z} S_m^{z} \rangle$ in Fig.~\ref{fig:USResultsXXZzz} as the input data.

After $d_z$ is obtained, the latent representation of all input data from previously trained AEs is taken with $n_{mid} = d_z$, and then do SA to estimate the optimal number of clusters $n$ via K-means. The
results are shown in (b) of Figs.~\ref{fig:USResultsXXZ2}, \ref{fig:USResultsXXZpm} and \ref{fig:USResultsXXZzz}, which specify that the mean $s$-score reaches the highest one when $n=3$ in  Figs.~\ref{fig:USResultsXXZ2} and \ref{fig:USResultsXXZzz}, however, for the input data as $\langle
S_0^{+} S_m^{-} \rangle$, $n=2$ in Fig.~\ref{fig:USResultsXXZpm}. This difference suggests that using $\langle S_0^{z} S_m^{z} \rangle$ as input data is possible to find the correct number of clusters (phases), whereas using $\langle S_0^{+} S_m^{-} \rangle$ can only recognize two clusters (phases) because they confuse FM and AFM phases. 

By projecting the multi-dimensional latent representations into a 2D space spanned by the first two principal components (features), we have (c) of Figs.~\ref{fig:USResultsXXZ2}, \ref{fig:USResultsXXZpm} and \ref{fig:USResultsXXZzz}. We first investigate Fig. ~\ref{fig:USResultsXXZ2} (c) for the input data as both
correlation functions. The plot suggests how the system could be divided into three clusters. Moreover, the points with blue color which represent the FM phase, are discontinuously separated from the other clusters. Such discontinuity is a signature of the first order phase transition at $\Delta_c =-1$. On the other hand, the green and orange lines are connected continuously, which represent a second order phase transition at $\Delta_c=1$. The same phenomena can be also seen in Fig.~\ref{fig:USResultsXXZzz}, whose data originate from $\langle S_0^{z} S_m^{z}\rangle$ correlations. 

However, for Fig.~\ref{fig:USResultsXXZpm} it looks quite different. In this figure $\langle S_0^{+} S_m^{-} \rangle$ correlations are used as input data, and they can not distinguish between FM and AFM phases. Therefore the latent representation does not have the signature of first order phase transition. The two curves are continuously connected. 

\begin{figure}[th] 
	\begin{center}
		\includegraphics[width=5cm]{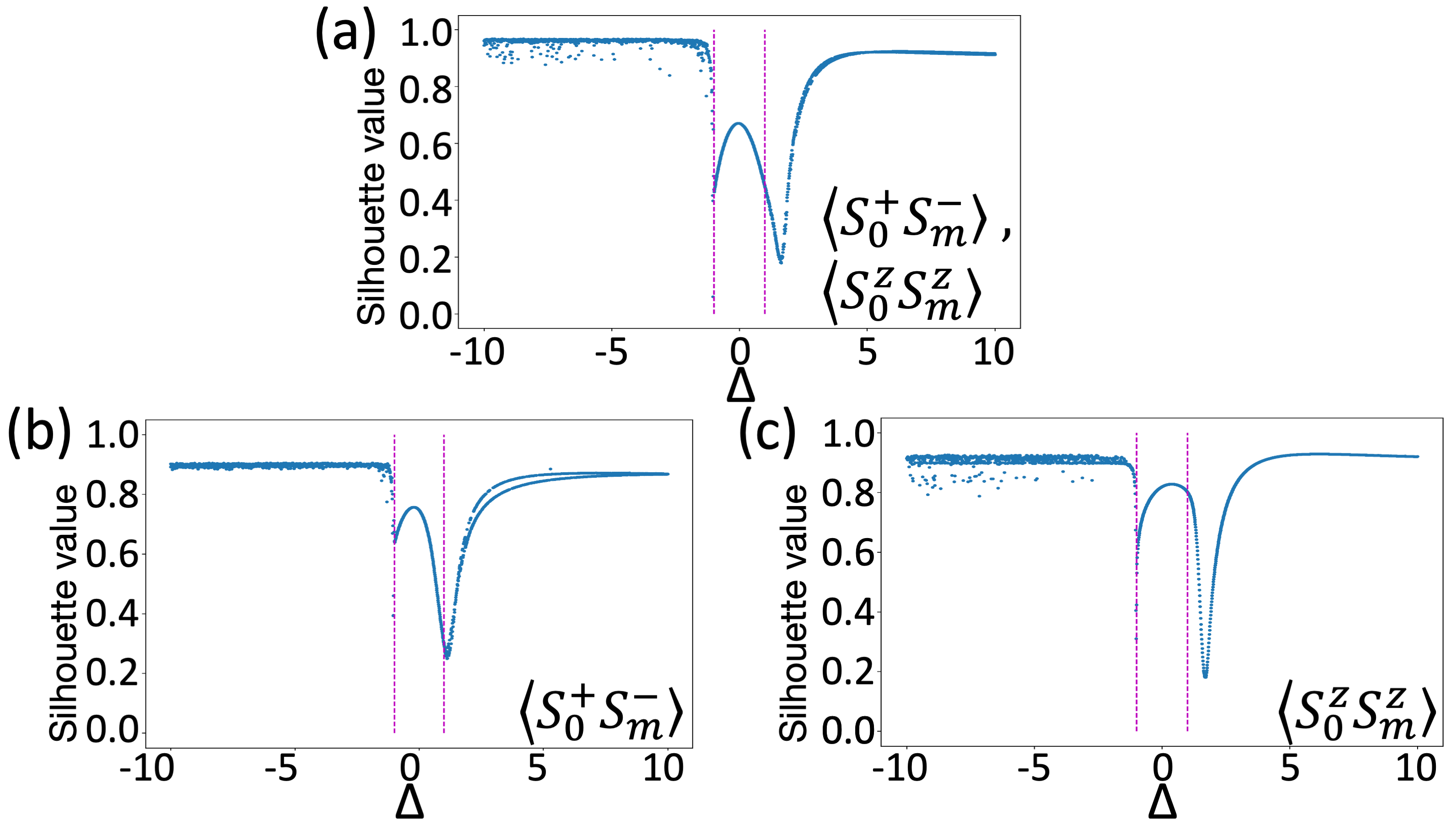}
		\caption{ Silhouette values as a function of $\Delta$ for $m_s=40$ with the input data as (a)
                  both $\langle S_0^{+} S_m^{-} \rangle$ and $\langle S_0^{z} S_m^{z} \rangle$, (b) $\langle
                  S_0^{+} S_m^{-} \rangle$, and (c) $\langle S_0^{z} S_m^{z} \rangle$. Some signatures are shown here: Silhouette values for the FM phase are almost the same, that means they all can be the most confident points. For the first-order phase transition, the minima of Silhouette values mostly lie around the critical point $\Delta=-1$, whereas for the second-order phase transition, they have some deviations from the critical point $\Delta=1$ except in the case  where the input data are $\langle S_0^{+}S_m^{-}\rangle$.}
		\label{fig:SAXXZ}
	\end{center}
\end{figure}

 Figs.~\ref{fig:SAXXZ}(a), (b) and (c) show Silhouette values of training data from both correlations, $\langle S_0^{+} S_m^{-} \rangle$, and  $\langle S_0^{z} S_m^{z} \rangle$, respectively. There are two minima of Silhouette values. On the left hand side of the figures, the minima are $-1.04$, $-1.05$ and $-1.01$ in Figs.~\ref{fig:SAXXZ} (a), (b) and (c), respectively, which lie almost at the critical points. However, those on the right hand side behave quite differently, where the minima lie at  $\Delta=1.41$, $1.12$ and $1.7$ in Figs.~\ref{fig:SAXXZ} (a), (b) and (c), respectively. For input data as $\langle S_0^{+} S_m^{-} \rangle$, the minimum is close to the critical point $1$, while for those as $\langle S_0^{z} S_m^{z} \rangle$ the minimum lies away from the TP. The reason for the latter case is due to the slow gap opening phenomenon for the second order phase transition, which means the correlation length is long near the critical point. When both correlation functions are stacked as input data, we obtain the median among the three minima shown in Figs.~\ref{fig:SAXXZ}(a), (b) and (c). As far as the maxima of Silhouette values are concerned, they are 6.08, 5.25, 6.3 for the AFM phase in Figs.~\ref{fig:USResultsXXZ2}, \ref{fig:USResultsXXZpm} and \ref{fig:USResultsXXZzz}, whereas for the FM phase, Silhouette values remain almost constant and the point of maximal Silhouette value changes for every new training, which means that almost every point represents the most confident one. Finally, the maxima of Silhouette values for the critical phase always occur at $\Delta=0$ no matter what kind of input data we feed.   

The TPs found by the input data as both correlation functions,  $\langle S_0^{+} S_m^{-} \rangle$ and $\langle S_0^{z}S_m^{z} \rangle$ are at the mean values $(-1.0280,1.6124)$, $(-1.0427, 1.1905)$ and $(-1.0144, 1.6849)$ with the standard deviations $(0.1446, 0.2007)$, $(0.0101, 0.2914)$ and $(0.0024, 0.0900)$, respectively, after collecting clustering results from 100 sets of latent representations via different trained AEs. For the first order phase transitions $\Delta_c=-1$, the predicted values are very precise
with small standard deviations, however, for the second order phase transition $\Delta_c=1$, the predicted results are bad with larger standard deviations. In order to make the predictions more precise, we further train a CNN classifier via supervised learning to predict all the TPs. To prepare the labeled data, we choose ten seeds in the following way: For the AFM phase, three seeds $(5.25, 6.08, 6.3)$ are chosen from the most confident points of SA in Figs.~\ref{fig:SAXXZ}(a), (b) and (c),  (see Discussion and Conclusion for more details). Instead, we choose the mirror reflecting points against 0, $(-5.25, -6.08,-6.3)$, as three seeds for the FM phase since each point in the FM phase has almost the same Silhouette value and thus each can be chosen deliberately as the most confident point. For the critical phase, we also choose three seeds $(-0.21, 0, 0.38)$, which are the most confident points of SA in Figs.~\ref{fig:SAXXZ} (a), (b) and (c).  Finally, an additional seed is chosen at $1.54$, which indicates the iDMRG cut-off point. We will have a  detailed explanation for these choices in Discussion and Conclusion. For each seed, we expand symmetrically around it by a window width $\Delta=0.1$ to obtain 1000 points with equal spacing. These 10000 labeled data points then form our training dataset, while the original 20001 ones become our test data set (no labels). As shown in (d) of Figs.~\ref{fig:USResultsXXZ2}, \ref{fig:USResultsXXZpm} and \ref{fig:USResultsXXZzz}, the phase boundaries obtained by CNN classifier are at mean values $(-1.0337, 1.1893)$, $(-1.0541,1.0218)$ and $(-1.0232, 1.2826)$ with standard deviations $(0.0050, 0.0309)$, $(0.0146, 0.0646)$ and $(0.0123, 0.0209)$. For the first order phase transition, the predicted values and standard deviations are similar to the case where no further supervision is applied, however, for the second order phase transition, the predicted values become more precise with smaller standard deviations. 

\section{Discussion and Conclusion} 
The proposed ML method by feeding the spin-spin correlation functions to predict the magnetic quantum phase transition points has shown its advantage with (supervised learning) or without (unsupervised learning) prior knowledge on phases of matter. However, there are still several issues we would like to mention here. 

\begin{figure}[th] 
	\begin{center}
		\includegraphics[width=9cm]{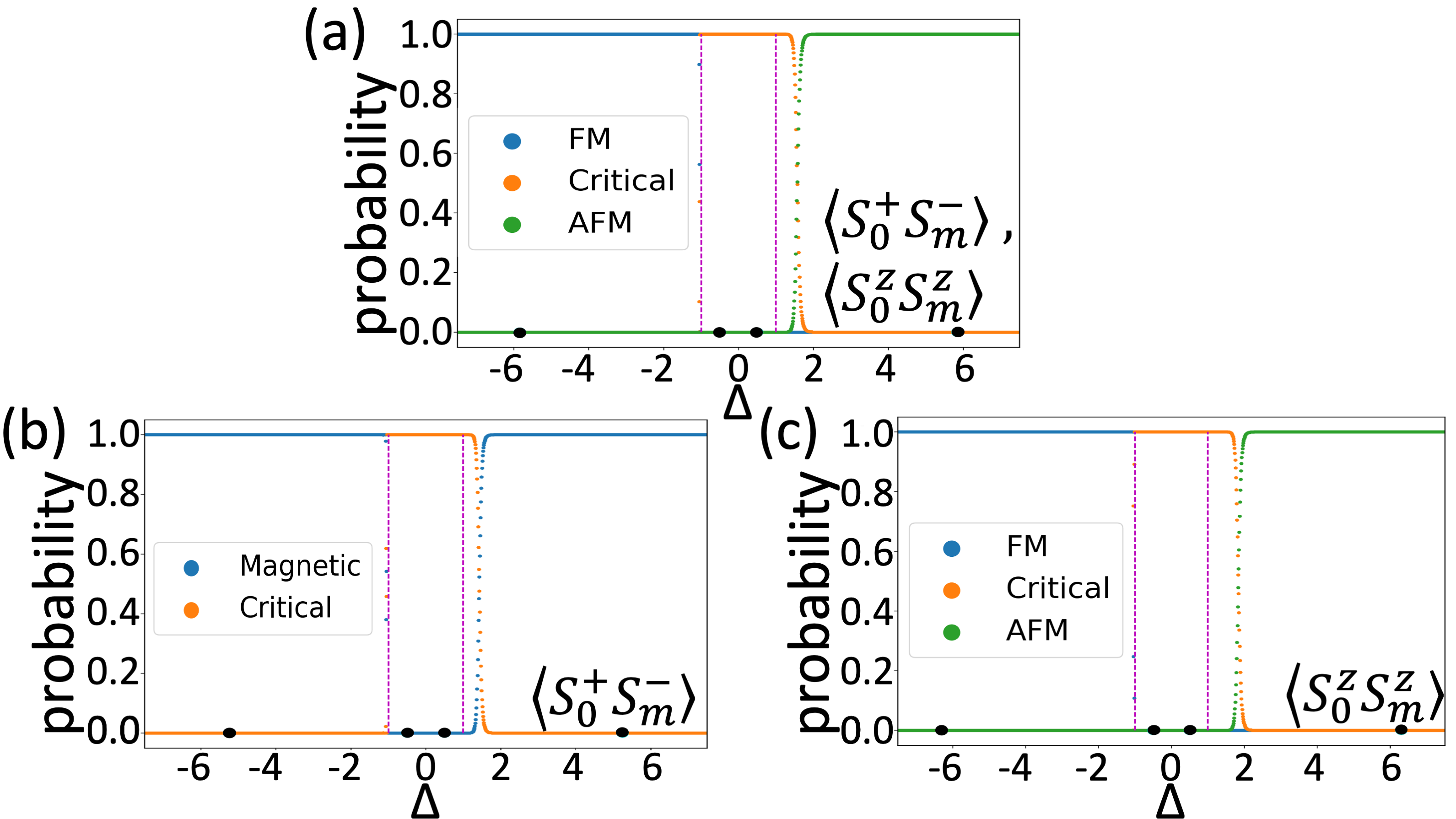}
		\caption{Supervised learning results with only four training seeds (gray dots) without the seeds related to the cut-off precision of iDMRG when the input data originate from (a) both $\langle S_0^{+} S_m^{-} \rangle$ and $\langle S_0^{z} S_m^{z} \rangle$, (b) $\langle S_0^{+} S_m^{-} \rangle$, and (c) $\langle S_0^{z} S_m^{z} \rangle$. For the first-order phase transition, the critical points remain intact, while for the second-order phase transition the results are worse than those with the extra training seeds related to the cut-off precision.  }
		\label{fig:4Points}
	\end{center}
\end{figure}

(1) In the supervised learning, we only used one training point (seed)  in each phase for the XY model to obtain precise phase transition points, however, for the XXZ model, we need two points. In order to show the necessity of using 6 training points, in Figs.~\ref{fig:4Points}(b) and (c) we show the trained results of the input data as $\langle S_0^{+} S_m^{-} \rangle$ and $\langle S_0^{z}S_m^{z} \rangle$, respectively, by using four training points, {\it i.e.,} one for the AFM phase, one for the FM phase and two for the critical phase for the sake of symmetry. The predicted phase transition points are at the mean values of $(-1.0577, 1.2700)$ and $(-1.0286, 1.7560)$ with the standard deviations $(0.0141, 0.0910)$ and $(0.0141, 0.0793)$ for the input data as $\langle S_0^{+} S_m^{-}\rangle$  and $\langle S_0^{z} S_m^{z} \rangle$, respectively. The
predicted TPs are on the same level with those of six training points for the first order phase transition point at $\Delta_c=-1$, however, they are less precise for the second order phase transition at $\Delta_c=1$. 

\begin{figure}[th] 
	\begin{center}
		\includegraphics[width=9cm]{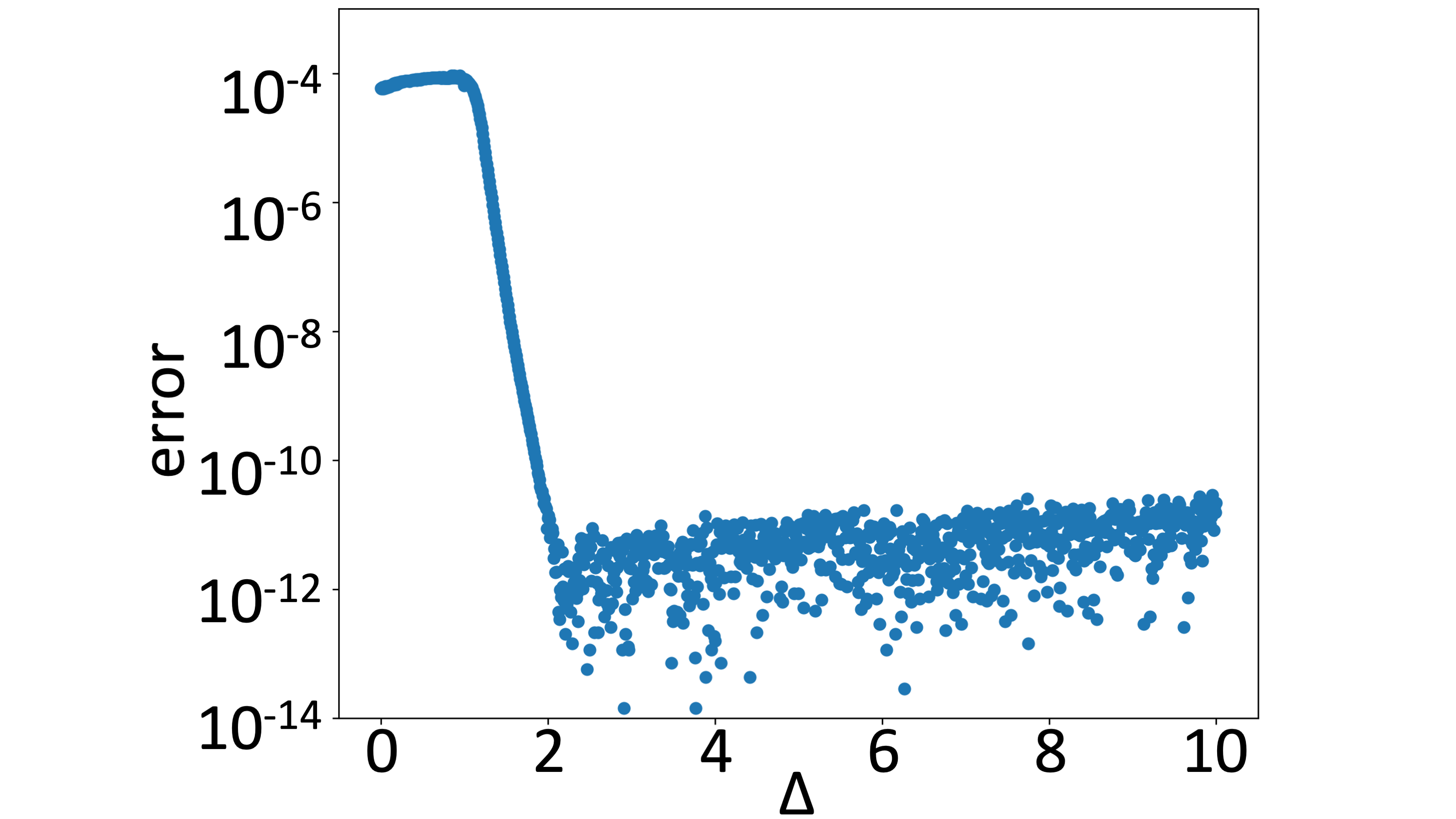}
		\caption{The errors of the ground states of the XXZ model with various positive $\Delta$ compared with the exact solution of Bethe Ansatz. }
		\label{fig:XXZAccuracy}
	\end{center}
\end{figure}

The basic reason behind this difference can be understood as follows. For the XY model, we can adjust the accuracy uniformly down to $10^{-16}$ by using Romberg integrations, however, for the XXZ model, the precision controlling is much more difficult. Fig.~\ref{fig:XXZAccuracy} shows the errors of the ground state energy by using iDMRG with the number of the bond dimension $m_b = 100$ compared with the exact solutions of Bethe Ansatz from $\Delta=0$ to $10$. In the critical regime, the errors are about $10^{-4}$ and then it drops quickly to $10^{-12}$ from $\Delta=1$ to $2$. In other words, the training data we calculate around the most confident point in SA has the precision up to $10^{-12}$, however, around the critical points the errors increase to $10^{-4}$. It turns out that the information around the most confident point can not sufficiently represent that around the critical point. Therefore, we need an extra training point (seed) near the second order phase transition point. After the calculation we found that $\Delta=1.536$ is just the point with a cut-off precision, $10^{-8}$, which can then be an appropriate extra seed around the critical TP. We have to check that the cut-off point we obtained is on the right hand side of the minimum within Silhouette values around $\Delta_c=1$ in Fig.~\ref{fig:SAXXZ}(a), because SA provides a suggestion where the approximated critical point could be. And this is just the case we met here. If it lies on the left hand side of the minimum, however, we should decrease our cut-off precision instead. 

\begin{figure}[th] 
	\begin{center}
		\includegraphics[width=9cm]{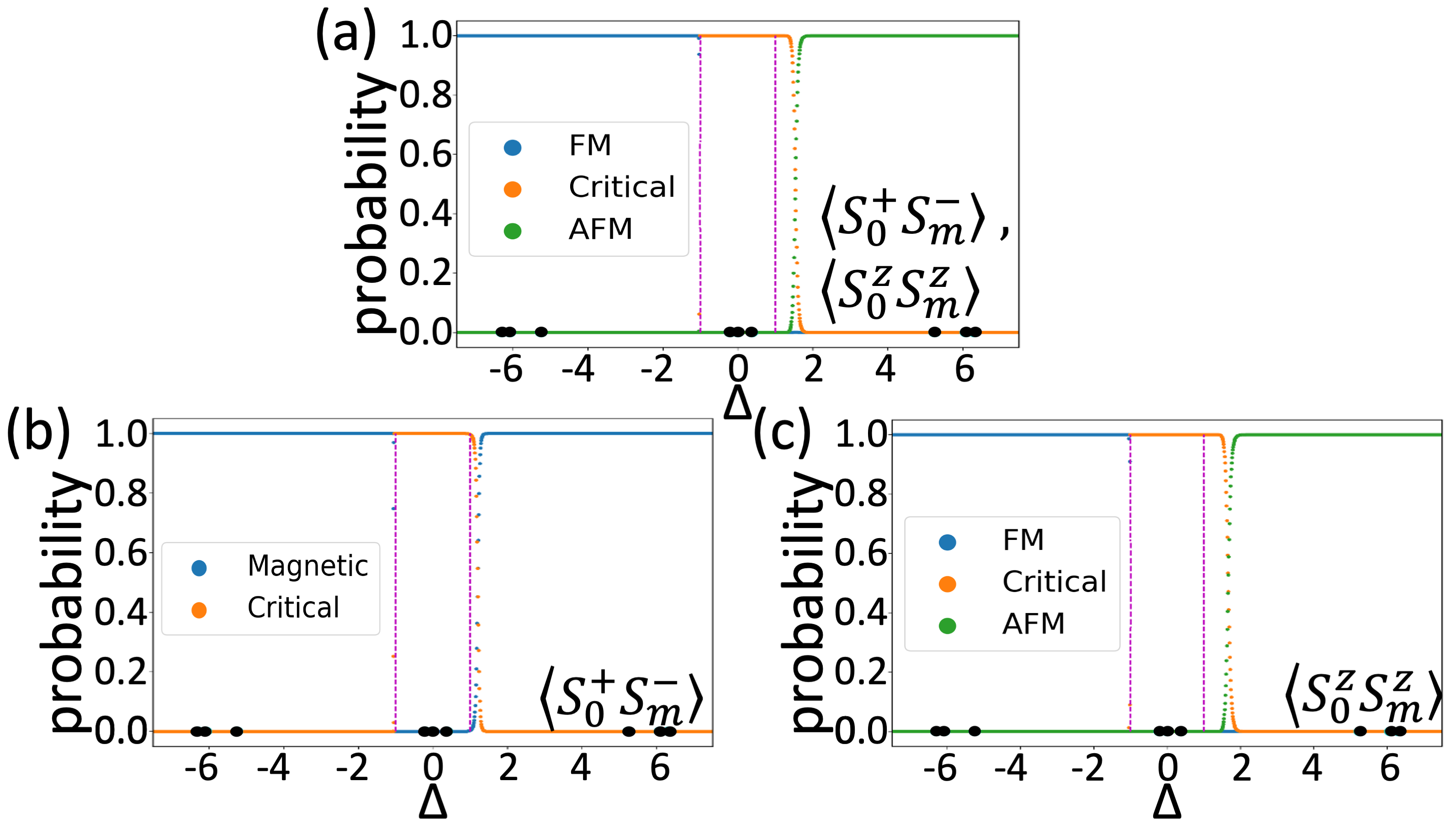}
		\caption{ In the last step of the unsupervised learning working protocol, we use supervised learning to improve precision of the predicted critical points. Here we show the supervised learning results with only nine training seeds (gray dots), including the most confident points of the three datasets using different correlation functions as the inputs, respectively. Note that the extra seed related to the numerical cut-off precision is not included. For the first-order phase transition, the critical points remain intact, while for the second-order phase transition the results are worse than those by using 9 training seeds plus the extra seed corresponding to the numerical cut-off precision (totally 10 seeds).}
		\label{fig:9Points}
	\end{center}
\end{figure}

In unsupervised learning we establish a way of finding training points (seeds) for the XXZ model. For the AFM and the critical phases, we choose three most confident points according to Silhouette values in the cases when the input data are both correlations, $\langle S_0^{+} S_m^{-}\rangle$ and $\langle S_0^{z} S_m^{z}\rangle$, respectively. In the FM phase we choose the mirror reflecting points of those in the AFM
phase. In Fig.~\ref{fig:9Points} we show the training results of those supervised learning using 9 training seeds. The TPs we obtain are at the mean values, $(-1.0339,1.5359)$, $(-1.0523,1.1751)$ and
$(-1.0238,1.7227)$, with the standard derivations, $(0.0061,0.0918)$, $(0.0155,0.1106)$ and $(0.0129,0.0823)$ corresponding to Figs.~\ref{fig:9Points}(a), (b) and (c), respectively. For easy comparison, both results with 10 and 9 training seeds are listed in Table.~\ref{tab:9or10}. We can see that
the results for the second order phase transitions are worse than those where we add an extra seed representing the cut-off property, whereas for the first order phase transition they remain not much changed. This suggests that we truly need the cut-off point to be a reference to fix the precision of the predicted phase boundary.  

\begin{table}[h]
        \centering
        \begin{tabular}{|l|c|c|c|c|c|l|}
            \hline
            & \multicolumn{2}{c|}{$\langle S_0^{z} S_m^{z} \rangle $, $\langle S_0^{+} S_m^{-}\rangle$}
            & \multicolumn{2}{c|}{$\langle S_0^{z} S_m^{z} \rangle$}
            & \multicolumn{2}{c|}{$\langle S_0^{+} S_m^{-} \rangle$}
            \\\hline
            & $\Delta_c^{1}$ & $\Delta_c^{2}$
            & $\Delta_c^{1}$ & $\Delta_c^{2}$
            & $\Delta_c^{1}$ & $\Delta_c^{2}$
            \\\hline
            9 seeds & -1.0339 & 1.5359 & -1.0238 & 1.7227 & -1.0523 & 1.1751 \\
            10 seeds & -1.0337 & 1.1893 & -1.0232 & 1.2826 & -1.0541 & 1.0218 \\
            \hline
       \end{tabular}
       \caption{Comparison of the predicted critical points obtained by using 9 and 10 training seeds.}
       \label{tab:9or10}
    \end{table}

 (2) For the XY model, $\rho_{xx}(m)$ for $m\approx \infty$ is significantly large in the PM phase, so is $\rho_{zz}(m)$ in the FM phase. They are defined as relevant correlations for the corresponding phases.  Similarly, for the XXZ model, $\langle S_0^{+} S_m^{-} \rangle$ is relevant in the critical phase only, whereas $\langle S_0^{z}S_m^{z} \rangle$ is relevant in both the FM and AFM phases. 

To distinguish whether correlations are relevant or irrelevant has its advantage. The relevant correlations can be used to find the phase boundaries between their corresponding phases and their neighbored phases. For instance, since $\rho_{xx}(m)$ is relevant for the PM phase, it can easily be used to find the TP
between PM and FM. Likewise $\rho_{zz}(m)$ can also be used to find the phase boundary. 

In contrast to the XY model, for the XXZ model the situation is more complicated. In unsupervised learning, $\langle S_0^{+}S_m^{-} \rangle$ can find two phase boundaries, one is that between the AFM and critical phases, and the other is between the critical and FM phases. However, by the clustering algorithm, they can only be grouped to two clusters. This is because the $\langle S_0^{+}S_m^{-} \rangle$ is only relevant for the critical phase, but irrelevant in both FM and AFM phases. Therefore, the machine can not distinguish the behaviors between the FM and AFM phases and only can differentiate the critical phase from the magnetic one.

On the other hand, there exists three clusters when $\langle S_0^{z}S_m^{z} \rangle$ correlations are used as input data. The reason is obvious: though $\langle S_0^{z}S_m^{z} \rangle$ are relevant for both FM and AFM phases, their behaviors are totally different. Therefore the machine is not confused between them. As a result, three phases are distinguished out of the training model, {\it i.e.}, two relevant phases and one irrelevant phase. 

Before further discussing about the XXZ model, one interesting point is worth mentioning here. Though $\rho_{yy}(m)$ is irrelevant for both PM and FM phases, however, by using this kind of input data, it can still find the phase boundary. The reason is that
the machine finds the different pattern of the correlations: In the PM phase, $\rho_{yy}(m)$ has a positive-negative oscillation for small $m$, while in the FM phase it remains positive. Therefore even the
correlations are irrelevant, they could sometimes to be used to distinguish the phases if the correlation patterns are different.   

(3) In the XXZ model there exists both first and second order phase transitions. This fact allows us to investigate the difference of the training processes and outcomes between them. 

First of all, the biggest difference lies on the feature maps (2D latent representation). In  Figs.~\ref{fig:USResultsXXZzz}(c), where the correlation functions $\langle S_0^{z}S_m^{z} \rangle$ are used as input data, the blue dots represent the points in the FM phase. It is obvious that those points are separated from the other orange (critical phase) and green (AFM phase) lines, which are continuously connected (in principle).  That means, the first order phase transition from the FM to critical phase is
characterized by two discontinuous clusters, whereas the second order phase transition is featured by two continuous lines. It is the analogy to the behavior of order parameters in the higher
dimensions. In the higher dimension (quantum criticality in dimensions higher than two, or classical criticality in dimensions higher than three), the local order parameter changes discontinuously across the first order critical point while continuously across the second order one. However, in 1D, there exist no symmetry breaking according to the Mermin-Wagner theorem, and thus the local order parameters are all zero in all phases. One can use the correlation functions instead of the local order parameters to find the phase transition points. By using the unsupervised (deep) learning, it is first found in this paper that the cluster points of the feature maps (2D latent representation) represent the same characteristic like the local order parameters if the relevant correlation functions are used as input data. This is highly interesting and deserves further investigations in the near future. 

Additionally, we observe that by using supervised learning or unsupervised learning, the first order phase transition is relatively easier to predict, whereas to find the second order phase transition is difficult when correlation functions are used as input data. It is because that the gaps in the FM phase are all similarly large with various $\Delta$, and suddenly drop to zero across the phase boundary to the other phase. For the machine, it is easily trained to distinguish between their differences. However, for the second order phase transition, the gap opens very slowly. Therefore the correlation functions are all similar across the phase boundary. Especially when the results are not so precise, the machine can be confused very easily, resulting in large error of the predicted second order phase boundary. In particular, that is also why we need an extra training seed related to the cut-off point of iDMRG to improve the precision of the predicted second order TP.  

(4) The Silhouette values also reveal some special features for the phase transitions. For instance, in the XXZ model Fig.~\ref{fig:SAXXZ} shows the Silhouette values of the data in the FM phase are almost the same, {\it i.e.,} all of them can be the most confident points. On the other hand, for the critical and AFM phases, Silhouette values have a wide range of difference.

For the XY model, $h=2$ is a second order phase transition from the PM to FM phase, whose behavior, as shown in Fig.~\ref{fig:SAXY}, is similar to that of the second order phase transition at $\Delta=1$ in the XXZ model (see Fig.~\ref{fig:SAXXZ}). Therefore we can infer that for the second order phase transition, because of the slowly gap opening, Silhouette value climbs from a small value to a constant one. This would be in sharp contrast to the case of the first order phase transition. Since the gap here closes very abruptly, those gapped states in the FM phase would have almost the same Silhouette value.    

Another interesting feature of SA is that the minima of Silhouette values are not far away from the phase transition points. For the XXZ model, the minimum of Silhouette values near the FM to the critical phase lies almost exactly at the critical point $\Delta=1$, whereas for the second order phase transition, the minima, which are at $\Delta=1.41, 1.12$ and $1.7$ in Figs.~\ref{fig:SAXXZ}(a), (b) and (c), respectively, do not exactly lie at the critical points. This also characterizes the difference between the first and second order phase transitions. 

To conclude, we use spin-spin correlation functions as input data to feed into machines in either a supervised or an unsupervised way of ML to find the phase transition points of the XY and XXZ models. The results show sharp boundaries and good precision to the exact values. Particularly in the unsupervised learning, the obtained latent representation (feature map) after dimension reduction and Silhouette values, which we have proposed in our previous work, provide insightful signatures of first order and second order phase transitions. We also show the importance of relevant correlation functions, which can be used as input data to find the phase boundaries between their corresponding phases and the neighboured ones. This concept is quite useful when the systems are more complicated and the phase boundaries have no exact solutions. What we have to do in such systems is to find out their relevant correlation functions and hence use them to find the phase boundaries. 

\section{Acknowledgement}
M.C.Chung acknowledges the NSTC  support under the contract Nos 111-2112-M-005 -013 - and the Asian Office of Aerospace Research and Development (AOARD) for support under Award No. FA2386-20-1-4049. I.P.M. acknowledges support from
the Australian Research Council (ARC) Discovery Project Grant No.
DP200103760.

\appendix
\section{Correlation Matrices of XY Models}

To obtain $\rho_{xx}$, $\rho_{yy}$ and $\rho_{zz}$ we follow Lieb, Schultz and Mattis\cite{LiebSchultzMattis}. In fact, 
\begin{equation}
\rho_{xx}(l-m) = \langle \Psi_0| (\sigma_l^{+} + \sigma_l^{-})
(\sigma_m^{+} + \sigma_m^{-}) |\Psi_m\rangle 
\end{equation}
Through the Jordan-Wigner transformation, Eq.~(\ref{JW}), we obtain
\begin{equation}
\rho_{xx}(l-m) = \langle \Psi_0| (c_l^{\dagger} - c_l) \exp{\left(i\pi \sum_{k=l+1}^{m-1}
                   c_k^{\dagger}c_k\right)}  (c_m^{\dagger} + c_m)
               |\Psi_0\rangle. 
\end{equation}
By using the fact that 
\begin{equation}
\exp{(i\pi c_k^{\dagger}c_k )} = (c_k^{\dagger} + c_k)(c_k^{\dagger}
-c_k),
\end{equation} 
we end up with the equation:
\begin{equation}
\rho_{xx} (l-m) = \langle \Psi_0| B_l A_{l+1} B_{l+1} \cdots A_{m-1}
B_{m-1} A_m    |\Psi_0\rangle,
\end{equation}
where $A_k = c_k^{\dagger} +c_k$ and $B_k= c_k^{\dagger} -c_k$. We hence perform the Wick's theorem\cite{Wick50} to obtain $\rho_{xx} $ in Eq.~(\ref{rhoxx}).

It is also similar for $\rho_{yy}$. $\rho_{yy}(l-m)$ has the form as follows:
\begin{equation}
  \rho_{yy} = \langle \Psi_0| (\sigma_l^{+} - \sigma_l^{-})
(\sigma_m^{+} - \sigma_m^{-}) |\Psi_m\rangle,
\end{equation}
which can be Jordan-Wigner transformed to 
\begin{equation}
\rho_{yy}(l-m) = (-1)^{l-m} \langle \Psi_0|  A_{l} B_{l+1} A_{l+1}\cdots B_{m-1}
A_{m-1} B_m    |\Psi_0\rangle,
\end{equation}
with the help of the equation:
\begin{equation}
 \exp{(i\pi c_k^{\dagger}c_k )} = -(c_k^{\dagger} - c_k)(c_k^{\dagger}
+c_k). 
\end{equation}
After we make use of the Wick's theorem, $\rho_{yy}$ is expressed as Eq.~(\ref{rhoyy}). 

Finally $\rho_{zz}(l-m)$ has the form:
\begin{equation}
\rho_{zz} =  \langle \Psi_0| (2\sigma_l^{+}\sigma_l^{-}-1)
(2\sigma_m^{+}  \sigma_m^{-}-1) |\Psi_m\rangle. 
\end{equation} 
In order to obtain $\rho_{zz}$, we have to use the identity,
\begin{eqnarray}
 2\sigma_k^{+}  \sigma_k^{-}-1 & = & -(\sigma_k^{+} +
 \sigma_k^{-})(\sigma_k^{+} - \sigma_k^{-}) \nonumber \\
& = & -(c_k^{\dagger} + c_k)(c_k^{\dagger} -c_k), 
\end{eqnarray}
and thus we have
\begin{equation}
  \rho_{zz}(l-m) = \langle \Psi_0| A_l B_l A_m B_m |\Psi_m\rangle. 
\end{equation}
After we make use of the Wick's theorem, the result becomes Eq.~(\ref{rhozz}).

To calculate $G_{l-m}$, we have to transform $H_{XY}$ into the momentum space as $H_{XY} - Nh/2 = \sum_k \mathbf{c_k^{\dagger}} H(k)\mathbf{c_k}$ with the notation, $\mathbf{c_k^{\dagger}} = (c_k,
c_{-k}^{\dagger})^T$, as Nambu particle-hole basis and 
\begin{equation}
H(k) = \mathbf{R} \cdot {\mathbf{\sigma}},
\end{equation}
where $\mathbf{R} = (0, -\gamma \sin{k}, \cos{k} + h/2)^T$ and $\mathbf{\sigma} = (\sigma_1, \sigma_2, \sigma_3)^T$ as a vector composed of three Pauli matrices. By using this notation, one can calculate first the correlation function in the momentum space,
\begin{equation}
G(k) \equiv \langle \mathbf{c_k} \mathbf{c_k}^{\dagger} \rangle, 
\end{equation}  
where $\langle \cdots \rangle$ denotes the expectation values of the ground state, $ \langle \Psi_0| \cdots |\Psi_0 \rangle $. After some straightforward algebra, we obtain
\begin{equation} \label{Gk} 
 G(k) = \frac{1}{2} \left(1+\frac{\mathbf{R} \cdot \sigma}{R} \right),
\end{equation} 
with the definition, $R \equiv |\mathbf{R}| = \sqrt{(\cos{k}+h/2)^2 + \gamma^2
  \sin^2{k}}$. Finally, through the relation,
\begin{equation}
\begin{split}
\langle B_l A_m\rangle & =  \frac{1}{L} \sum_k e^{-i k (r_l-r_m)}
\langle  c_k^{\dagger} c_{-k}^{\dagger}\rangle \\
 &- \langle  c_kc_{-k} \rangle - \langle  c_k c_k^{\dagger}\rangle +
 \langle  c_{-k}^{\dagger} c_{-k}   \rangle,
\end{split}
\end{equation}
and using Eq.(\ref{Gk}), we obtain Eq.~(\ref{BA}). 

\section{Model Architectures}
Here we present explicit neural network architectures for our numerical results shown in the subsection B of Sec. IV. and Sec. V. In the case of the XY model, for the AE used for Figs.~\ref{fig:USResultsXY3}, \ref{fig:USResultsXYxx}, \ref{fig:USResultsXYyy}, and \ref{fig:USResultsXYzz}, the model architecture is given in Table~\ref{tab:AEforXY}.

\begin{table}[h]
	\centering
	\begin{tabular}{lcc}
		\hline\hline
		Layer & Parameters & Activation
		\\\hline
		&  Input: $c \times 20 \times 20$  &  \\
		Conv. & $16 \times 3 \times 3$ & ReLU \\
		Linear & $c\times 6400 \times 512$ & ReLU \\
		Linear & $512 \times 256$ & ReLU \\
		Linear & $256 \times 128$ & ReLU \\
		Linear & $128 \times 64$ & ReLU \\
		Linear & $64 \times$ features & ReLU \\
		Linear & features $\times 64$ & ReLU \\
		Linear & $64 \times 128$ & ReLU \\
		Linear & $128 \times 256$ & ReLU \\
		Linear & $256 \times 512$ & ReLU \\
		Transposed conv. & $c \times 3 \times 3$ & Sigmoid \\
		\hline\hline
	\end{tabular}
	\caption{Model architecture of AE used for Figs.~\ref{fig:USResultsXY3}, \ref{fig:USResultsXYxx}, \ref{fig:USResultsXYyy}, and \ref{fig:USResultsXYzz}. Note that $c=3$ when using all $\rho_{xx}, \rho_{yy}, \rho_{zz}$ together as the input data, while $c=1$ when using $\rho_{xx}, \rho_{yy}$ or $\rho_{zz}$, respectively, as the input data. Moreover, ``features'' in the Parameters column indicates the number of neurons in the middlemost layer of AE, $n_{mid}$.}
	\label{tab:AEforXY}
\end{table}

As to fine-tuning the phase boundaries via supervised learning, additional CNN models are employed with the same architectures shown in Figs.~\ref{fig:CNN}(a) and (b).

On the other hand, in the case of the XXZ model, the architecture of AE used for Figs.~\ref{fig:USResultsXXZ2}, \ref{fig:USResultsXXZpm}, and \ref{fig:USResultsXXZzz} is given in Table~\ref{tab:AEforXXZ}.

\begin{table}[h]
	\centering
	\begin{tabular}{lcc}
		\hline\hline
		Layer & Parameters & Activation
		\\\hline
		&  Input: $c \times 40 \times 40$  &  \\
		Conv. & $16 \times 3 \times 3$ & ReLU \\
		Linear & $51200 \times 512$ & ReLU \\
		Linear & $512 \times 256$ & ReLU \\
		Linear & $256 \times 128$ & ReLU \\
		Linear & $128 \times 64$ & ReLU \\
		Linear & $64 \times$ features & ReLU \\
		Linear & features $\times 64$ & ReLU \\
		Linear & $64 \times 128$ & ReLU \\
		Linear & $128 \times 256$ & ReLU \\
		Linear & $256 \times 512$ & ReLU \\
		Transposed conv. & $c \times 3 \times 3$ & Sigmoid \\
		\hline\hline
	\end{tabular}
	\caption{Model architecture of AE used for Figs.~\ref{fig:USResultsXXZ2}, \ref{fig:USResultsXXZpm}, and \ref{fig:USResultsXXZzz}. Note that $c=2$ when using both $\langle S_0^{z} S_m^{z} \rangle$ and $\langle S_0^{+} S_m^{-}\rangle$ as the input data, while $c=1$ when using $\langle S_0^{z} S_m^{z} \rangle$ or $\langle S_0^{+} S_m^{-}\rangle$, respectively, as the input data. Moreover, ``features'' in the Parameters column indicates the number of neurons in the middlemost layer of AE, $n_{mid}$.}
	\label{tab:AEforXXZ}
\end{table}

Similarly, CNN models are employed to fine-tune the phase boundaries via supervised learning. They basically have the same architectures shown in Figs.~\ref{fig:CNN}(c) and (d), except that the number of neurons in the output layer, $n$, now should be determined by the outcome of K-means clustering analysis.

\end{document}